\documentclass{aa}  
\usepackage{graphicx}
\usepackage{txfonts}
\usepackage{hyperref}
\hypersetup{
    colorlinks=true,
    linkcolor=blue,
    filecolor=magenta,      
    urlcolor=cyan,
    citecolor=blue
    }
\usepackage{amsmath}
\usepackage[version=4]{mhchem}
\usepackage{gensymb}
\usepackage{chemfig}
\usepackage{comment}

\linenumbers
\renewcommand\makeLineNumber{}

\begin{document}

   \title{Beyond monoculture: Polydisperse moment methods for sub-stellar atmosphere cloud microphysics}

   \subtitle{II. A three-moment gamma distribution formulation for GCM applications}

   \author{Elspeth K.H. Lee\inst{1} and Kazumasa Ohno\inst{2}}

   \institute{$^{1}$Center for Space and Habitability, University of Bern, Gesellschaftsstrasse 6, CH-3012 Bern, Switzerland \\ 
   $^{2}$Division of Science, National Astronomical Observatory of Japan, 2-21-1 Osawa, Mitaka-shi, Tokyo, Japan }

   \date{Received DD MM YYYY / Accepted DD MM YYYY}
 
  \abstract
   {Understanding how the shape of cloud particle size distributions affects the atmospheric properties of sub-stellar atmospheres is a key area to explore, particularly in the JWST era of broad wavelength coverage, where observations are sensitive to particle size distributions. 
   It is therefore important to elucidate how underlying cloud microphysical processes influence the size distribution, in order to better understand how clouds affect observed atmospheric properties.}
   {In this follow-up paper, we aim to extend our sub-stellar atmosphere microphysical cloud formation framework to include effects of assuming a polydisperse gamma particle size distribution, requiring a three-moment solution set of equations.}
   {We develop a three-moment framework for sub-stellar mineral cloud particle microphysical nucleation, condensation, evaporation and collisional growth assuming a gamma distribution.
   As in our previous work, we demonstrate the effects of polydispersity using a simple one-dimensional Y-dwarf KCl cloud formation scenario, and compare the results with the monodisperse case.}
   {Our three-moment scheme provides a generalised framework applicable to any size distribution with a defined moment generation expression.
   In our test case, we show that the gamma distribution evolves with altitude, initially broad at the cloud base and narrowing at lower pressures.
   We find that differences between the gamma and monodisperse cloud structures can be significant, depending on the surface gravity of the atmosphere.}
   {We present a self-consistent framework for including the effects of polydispersity for sub-stellar microphysical cloud studies using the moment method.}


   \authorrunning{Lee \& Ohno}
   \maketitle

\section{Introduction}

The size distribution of cloud particles plays a critical role in shaping the observable properties of sub-stellar atmospheres and significantly influences the thermal balance through scattering and absorption opacities, which depend on the composition of the condensing material.
Recent evidence for complex particle size distributions in sub-stellar atmospheres includes the inference of nanometre-sized particles in WASP-17b \citep{Grant2023}, consistent with a solid \ce{SiO2} composition based on infrared absorption features. 
Retrieval modelling of brown dwarf atmospheres \citep{Burningham2021} also suggests a cloud profile comprising various particle sizes, with $\sim$0.1 $\mu$m silicate particles being dominant and lofted to high altitudes.

Typically, microphysical models of cloud formation in sub-stellar atmospheres use equation sets that assume a monodisperse distribution, utilising a single representative average value \citep[e.g.][]{Woitke2003, Helling2008, Ohno2018, Lee2023}.
However, the monodisperse formulation neglects the effects of polydispersity on cloud formation processes such as condensation, evaporation, and collisional growth, where the variety of particle sizes can significantly influence the overall cloud structure.
Simulations using bin-resolving models such as the CARMA model \citep[e.g.][]{Gao2018, Powell2019} applied to hot Jupiter and brown dwarf atmospheres suggest a range of particle sizes can be present in sub-stellar atmospheres, as well as potential multi-modality.
These results highlight the importance of accounting for polydispersity when modelling cloud formation in sub-stellar atmospheres.

Moment (or bulk) cloud microphysical methods offer a useful middle ground between the more computationally expensive bin (or spectral) models \citep[e.g.][]{Gao2018, Powell2024} and simplified schemes such as tracer saturation adjustment \citep[e.g.][]{Tan2019, Komacek2022}. 
Consequently, the moment method is frequently coupled to large-scale 3D atmospheric simulations, such as General Circulation Models (GCMs), in the Earth science community \citep[e.g.][]{Morrison2008}, hot Jupiter exoplanets \citep[e.g.][]{Lee2016, Lines2018, Lee2023} and brown dwarf atmospheres \citep[e.g.][]{Lee2025}.
However, to date, only monodisperse moment methods have been employed in GCM simulations. 
While these can predict a representative particle size, they cannot self-consistently predict the width of the size distribution without additional assumptions.
Other schemes use additional assumptions on the properties of the size distribution and particle sizes, such as time-independent equilibrium balance between vertical transport and settling velocity \citep{Ackerman2001}, or localised, instantaneous formation/rain out \citep[e.g.][]{Parmentier2016,Roman2019}.
These approaches generally do not evolve a tracer within the GCM, but instead diagnose cloud properties directly from the local temperature and chemical structure of each vertical profile.
Of particular relevance to the current study is the work of \citet{Christie2022}, who derived a variant of the \citet{Ackerman2001} model that assumes a gamma distribution rather than the commonly used log-normal.

To overcome this limitation, several studies in the Earth cloud science adopt the three-moment method that uses the shape of the size distribution as a prognostic variable \citep[e.g.,][]{Milbrandt&Yau05_3moment,Naumann&Seifert16,Paukert+19}.
In this follow up study to \citet{Lee2025} and \citet{Lee2025p} (Paper I), which presented two-moment methods with a monodisperse and exponential size distribution respectively, we extend the model to a three-moment method that is capable of predicting the evolution of size distribution shape. 
Due to the uncertainty in the exact size distribution properties present in sub-stellar atmospheres, we attempt to derive size distribution properties self-consistently with minimal parameterisations.
As often assumed in Earth cloud models, we adopt the gamma particle size distributions to describe the underlying size distribution of cloud particles.
In Section \ref{sec:gamma_motivation}, we introduce the observational and theoretical motivation to assume the gamma size distribution.
In Section \ref{sec:gamma_dist}, we present the properties of the gamma distribution, its moments and introduce the three-moment microphysical scheme.
In Section \ref{sec:gamma_cond}, we derive the condensation and evaporation rates for the gamma distribution.
In Section \ref{sec:gamma_coll}, we derive collisional growth rates, taking into account the polydispersity introduced by the gamma distributions.
In Section \ref{sec:1D}, we apply our new methodology to the same Y-dwarf KCl cloud formation case as in Paper I.
Section \ref{sec:disc} and Section \ref{sec:conc} contain the discussion and conclusions respectively of our study.

\section{Introduction of gamma distribution}
\label{sec:gamma_motivation}

\begin{figure}
    \centering
    \includegraphics[width=\linewidth]{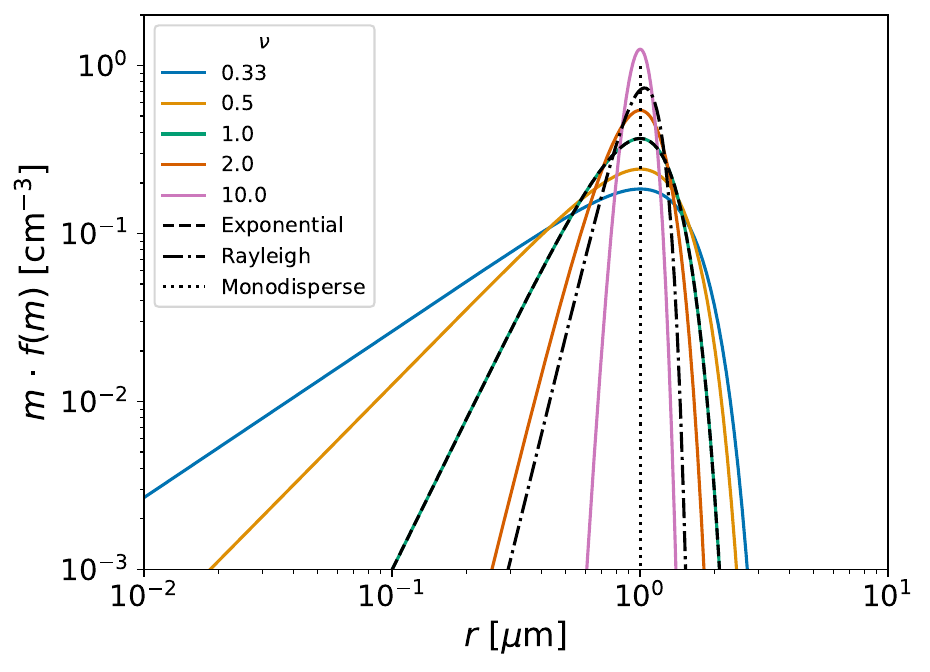}
    \caption{Gamma size distribution compared to the exponential, Rayleigh and monodisperse distributions for various values of $\nu$. 
    The representative particle size is $r_{\rm c}$ = 1 $\mu$m and total number density is $N_{\rm c}$ = 1 cm$^{-3}$.}
    \label{fig:gamma_dist}
\end{figure}

The gamma particle mass distribution, $f(m)$ [cm$^{-3}$ g$^{-1}$], is given by
\begin{equation}
\label{eq:exp_dist}
    f(m) = \frac{N_{\rm c}}{\lambda^{\nu}\Gamma(\nu)}m^{\nu - 1}e^{-m/\lambda},
\end{equation}
where $N_{\rm c}$ [cm$^{-3}$] is the total number density of the distribution, $\lambda$ [g] is the scale parameter, $\nu$ the shape parameter, $\Gamma(z)\equiv \int_{\rm 0}^{\infty}t^{z-1}e^{-t}dt$ is the gamma function, and $m$ [g] is the mass of an individual particle.
When $\nu$ = 1, the gamma distribution reduces to an exponential distribution, which was explored in Paper I.
$\lambda$ and $\nu$ are related to the mean mass (expectation value E[$m$] [g]) and variance (Var[$m$] [g$^{2}$]) of the distribution through
\begin{equation}
    {\rm E}[m] = \lambda\nu,
\end{equation}
and
\begin{equation}
    {\rm Var}[m] = \lambda^{2}\nu,
\end{equation}
respectively.
In Figure \ref{fig:gamma_dist}, we illustrate how the gamma distribution changes with shape parameter $\nu$. 
In general, the gamma distribution consists of a power-law tail at smaller particle sizes with an exponential cut off at large sizes.
As $\nu$ $\rightarrow$ 0, the distribution widens, becoming more polydisperse, while as $\nu$ $\rightarrow$ $\infty$ the distribution narrows, showing more monodisperse characteristics.

\subsection{Observational motivation}

There are several observational motivations for adopting the gamma distribution.
A seminal work of \citet{Marshall1948} showed that the size distribution of rain droplets in the Earth's atmosphere can be well fitted by an exponential function, commonly known as the Marshall-Palmer distribution\footnote{The Marshall-Palmer distribution is given by 
\begin{equation}
dn/dD=N_{\rm 0}\exp{(-\lambda_{\rm D}D)},
\end{equation}
where $D$ is particle diameter.
Note that Earth cloud studies often describe the size distribution as a number density per radius, not the number density per mass adopted in this study.}, except at small particle sizes.
Later studies showed that the gamma distribution is sufficiently versatile to fit the observed cloud properties \citep[e.g.,][]{Ulbrich83,McFarquhar+15}.
Beyond Earth clouds, \citet{Hansen&Hovenier74} showed that sulfuric acid clouds modelled with the Hansen distribution\footnote{
The Hansen distribution is given by
\begin{equation}
    dn/dr=r^{(1-3b)/b}\exp{(-r/ab)},
\end{equation}
where $C$ is a constant, $r$ is the particle radius, $a$ is the area-weighted averaged radius, and $b$ is the effective variance.}, a variant of the gamma distribution, can reproduce the observed polarisation signatures of Venusian clouds.
In brown dwarf atmospheres, \citet{Hiranaka+16} showed that the gamma distribution containing abundant sub-micron particles can well explain the near-infrared colours of some red L dwarfs.
Retrieval modelling by \citet{Burningham2021} also found that a gamma distribution is strongly preferred over a log-normal distribution when fitting the emission spectrum of the red L dwarf 2M2224-0158.

\subsection{Physical motivation}

In addition to observational evidence, there are several physical motivations for adopting the gamma distribution.
The fundamental equation describing the evolution of particle size distribution is given by \citep[see, e.g., Appendix A of][]{Lee2025}
\begin{equation}\label{eq:SD_evolve}
    \frac{\partial f(m)}{\partial t}=\frac{\partial}{\partial z}\left[\rho_{\rm a}K_{\rm zz}\frac{\partial}{\partial z}\left( \frac{f(m)}{\rho_{\rm a}}\right)-v_{\rm f}(m)f(m) \right]-\frac{\partial}{\partial m}\left[\left(\frac{dm}{dt}\right)f(m)\right],
\end{equation}
where $z$ [cm] is the altitude, $\rho_{\rm a}$ [g cm$^{-3}$] is the atmospheric density, $v_{\rm f}$ [cm s$^{-1}$] is the terminal settling velocity, $(dm/dt)$ [g s$^{-1}$] the rate of change in particle mass due to microphysical processes such as condensation and $K_{\rm zz}$ [cm$^{2}$ s$^{-1}$], the eddy diffusion coefficient. 
For illustrative purposes here, we ignore collisional growth and adopt a 1D framework that approximates vertical transport by atmospheric circulation as an eddy diffusion term.

To gain physical insights into the gamma distribution, we conduct an order-of-magnitude analysis of Eq. \eqref{eq:SD_evolve}.
Focusing on a single vertical layer, the vertical transport sink term acts to remove cloud particles from the layer, while the second term acts to increase or decrease the mass of the particles inside the layer.
Thus, one may crudely approximate $\partial/\partial z\sim 1/H_{\rm a}$, where $H_{\rm a}$ [cm] is the atmospheric altitude length scale, to rewrite the equation as
\begin{equation}
    \frac{\partial}{\partial m}\left[\left(\frac{dm}{dt}\right)f(m)\right]\sim -\left(\frac{K_{\rm zz}}{H_{\rm a}^2}+\frac{v_{\rm f}(m)}{H_{\rm a}}\right)f(m),
\end{equation}
where we have considered a steady state.
As shown later, the mass growth rate and settling velocity can be often expressed as a power-law function of mass.
Let $(dm/dt)=Am^{\alpha}$ and $v_{\rm f}(m)=Bm^{\beta}$, we can further rewrite the equation as
\begin{equation}
    \frac{\partial}{\partial m}\left[ m^{\alpha}f(m)\right]\sim -\frac{1}{Am^{\alpha}}\left(\frac{K_{\rm zz}}{H_{\rm a}^2}+\frac{Bm^{\beta}}{H_{\rm a}}\right)m^{\alpha}f(m).
\end{equation}
One can analytically solve this equation to obtain the steady-state size distribution as
\begin{equation}
    f(m)\propto m^{-\alpha}\exp{\left[-\left( \frac{K_{\rm zz}}{(1-\alpha)H_{\rm a}^2}+\frac{Bm^{\beta}}{(1+\beta-\alpha)H_{\rm a}}\right)\frac{m^{1-\alpha}}{A}\right]}.
\end{equation}
The obtained analytic distribution consists of a power-law tail at small particle mass with an exponential cutoff at large mass, reproducing the salient characteristics of the gamma distribution.
This analysis indicates that the power-law tail at low particle mass encapsulates the information of mass dependence of particle growth.
By introducing a particle mass growth timescale $\tau_{\rm grow}=m/(dm/dt)=m^{1-\alpha}/A$ [s], an eddy diffusion timescale $\tau_{\rm edd}=H_{\rm a}^2/K_{\rm zz}$ [s], and a settling timescale $\tau_{\rm set}=H_{\rm a}/v_{\rm f}=H_{\rm a}/Bm^{\beta}$ [s], one can re-write the analytic distribution into a physically informative form
\begin{equation}
    f(m)\propto m^{-\alpha}\exp{\left[-\left( \frac{\tau_{\rm grow}}{(1-\alpha)\tau_{\rm edd}}+\frac{\tau_{\rm grow}}{(1+\beta-\alpha)\tau_{\rm set}}\right)\right]}.
\end{equation}
Thus, the distribution follows a power-law behavior at masses at which the growth is faster than vertical transport, i.e., $\tau_{\rm grow}\ll \tau_{\rm edd}$ and $\tau_{\rm grow}\ll \tau_{\rm set}$, whereas the distribution exponentially drops once the vertical transport is faster than the growth.
Note that \citet{Garrett19} presented a similar analysis to provide a physical reason why rain droplets in the Earth closely follow the gamma distribution.
Although the derivation presented above is potentially overly simplistic to accurately predict the real size distribution, it provides a natural physical motivation to adopt the gamma distribution when developing a polydisperse particle distribution model including cloud microphysics.

\subsection{Moments with assumed gamma distribution}
\label{sec:gamma_dist}

We present a derivation of the mass moments assuming a gamma distribution, along with the source terms required for evolving the cloud microphysics ordinary differential equation (ODE) system in time.
The mass moments of the particle mass distribution, $M^{(k)}$ [g$^{k}$ cm$^{-3}$], are given by
\begin{equation}
    M^{(k)} = \int_{0}^{\infty}m^{k}f(m)dm,
\end{equation}
where $k$ is the order of the moment, which can take an integer or non-integer value.
Inserting the gamma distribution in the moment definition equation gives
\begin{equation}
    M^{(k)} = \frac{N_{\rm c}}{\lambda^{\nu}\Gamma(\nu)}\int_{0}^{\infty}m^{k + \nu - 1}e^{-m/\lambda}dm.
\end{equation}
After some algebra and applying the definition of the gamma integral, the moment generator for the gamma mass distribution is analytically expressed as
\begin{equation}
\label{eq:mom_gen}
    M^{(k)} = N_{\rm c}\lambda^{k}\frac{\Gamma(\nu + k)}{\Gamma(\nu)}.
\end{equation}

Each integer moment represents bulk values of the particle size distribution, such as the total number density, $N_{\rm c}$ [cm$^{-3}$], for $k$ = 0
\begin{equation}
    M^{(0)} = N_{\rm c},
\end{equation}
the $k$ = 1 moment that represents the total mass density, $\rho_{\rm c}$ [g cm$^{-3}$], of the cloud particles
\begin{equation}
  M^{(1)} = \rho_{\rm c},
\end{equation}
and the $k$ = 2, $Z_{\rm c}$ [g$^{2}$ cm$^{-3}$], moment
\begin{equation}
  M^{(2)} = Z_{\rm c},
\end{equation}
that is related to the Rayleigh scattering cross-section of the particles.

The parameters $\lambda$ and $\nu$ can be estimated from the computed integer moments. 
The number-weighted mean mass, $m_{\rm c}$ [g], is given by the ratio of the first to zeroth moment:
\begin{equation}
    m_{\rm c} = \frac{\rho_{\rm c}}{ N_{\rm c}} = \lambda\nu.
\end{equation}
The variance, $\sigma_{\rm c}^{2}$ [g$^{2}$], is given by
\begin{equation}
  \sigma_{\rm c}^{2} = \frac{Z_{\rm c}}{N_{\rm c}} - \left(\frac{\rho_{\rm c}}{N_{\rm c}}\right)^{2} = \lambda^{2}\nu.
\end{equation}
Rearranging gives expressions for the shape and scale parameters
\begin{equation}
    \nu = \frac{m_{\rm c}^{2}}{\sigma_{\rm c}^{2}},
\end{equation}
and
\begin{equation}
\label{eq:lambda}
    \lambda = \frac{m_{\rm c}}{\nu},
\end{equation}
respectively.
The representative particle radius, $r_{\rm c}$ [cm], corresponding to the number-weighted mean mass, $m_{\rm c}$, is given by
\begin{equation}
\label{eq:rc}
    r_{\rm c} = \left(\frac{3m_{\rm c}}{4\pi\rho_{\rm d}}\right)^{1/3},
\end{equation}
where $\rho_{\rm d}$ [g cm$^{-3}$] is the particle bulk density.
Hereafter, we refer to $r_{\rm c}$ as the ``\textit{representative radius}''.
Similarly, the mass-weighted mean mass can be defined as
\begin{equation}
    m_{\rm p} =\frac{\int_{\rm 0}^{\rm \infty} m{\cdot}mf(m)dm}{\int_{\rm 0}^{\rm \infty} mf(m)dm}= \frac{M^{(2)}}{M^{(1)}}=\lambda(\nu+1).
\end{equation}
This expression highlights that the number-weighted mean mass $m_{\rm c}$ is representative in the $\nu$ $\gg$ 1 regime, whereas the mass-weighted mean mass $m_{\rm p}$ deviates from $m_{\rm c}$ for $\nu$ $\ll$ 1, where the distribution becomes broader and polydispersity plays a more significant role.
The representative radius of the mass-dominating particles, $r_{\rm p}$ [cm], is given by
\begin{equation}
\label{eq:rp}
    r_{\rm p} = \left(\frac{3m_{\rm p}}{4\pi\rho_{\rm d}}\right)^{1/3}=r_{\rm c}\left(\frac{\nu+1}{\nu}\right)^{1/3}.
\end{equation}
In this study, we consider only spherical, homogeneous cloud particles, allowing Eqs. \eqref{eq:rc} and \eqref{eq:rp} to be applied. Porous particles and fractal geometries, as considered in some previous studies \citep[e.g.][]{Ohno2020,Samra+20}, are not included.

\section{Condensation with gamma distribution}
\label{sec:gamma_cond}

\begin{figure}
    \centering
    \includegraphics[width=0.95\linewidth]{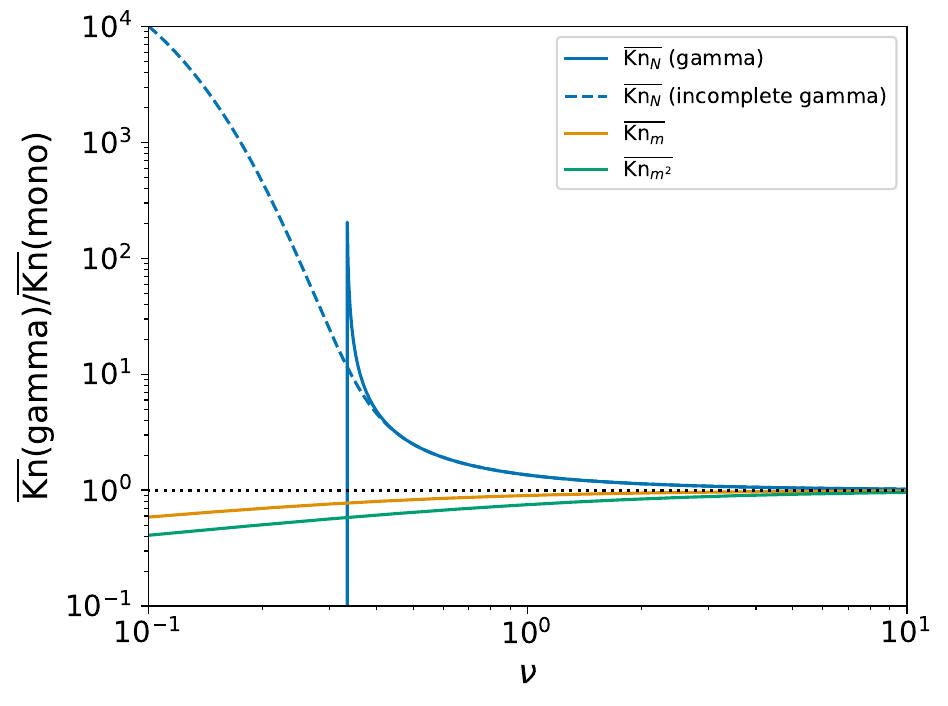}
    \caption{Ratio of the population averaged Knudsen numbers with the monodisperse Knudsen number.
    Due to the gamma function properties, a pole occurs at values of $\nu$ $\leq$ 1/3 for the number-weighted Knudsen number using Eq. \eqref{eq:Kn_N} (blue solid line), leading to singularities at $\nu$ $\leq$ 1/3.
    The incomplete gamma function formulation using Eq. \eqref{eq:Kn_N_fix}, which accounts for a cutoff at small particle sizes, avoids the singularities in the gamma function (dashed blue line) and is valid across the $\nu$ range.}
    \label{fig:Kn_pop}
\end{figure}

In this section, we derive the condensation and evaporation rates for a cloud particle population described by a gamma distribution. 
Unlike the exponential case (Paper I), the gamma formulation requires additional expressions involving the second moment, in addition to the first.
In the method of moments \citep[e.g.][]{Gail2013}, the rate of change of the $k$-th moment due to condensation or evaporation is given by
\begin{equation}
\label{eq:dMk_dt}
    \frac{dM^{(k)}}{dt} = k\int_{0}^{\infty}m^{k-1}\frac{dm}{dt}f(m)dm,
\end{equation}
where $dm/dt$ [g s$^{-1}$] is the rate of change of the cloud particle mass.
The particle Knudsen number, Kn, is defined as
\begin{equation}
    {\rm Kn} = \frac{\lambda_{\rm a}}{r},
\end{equation}
where $r$ [cm] is the radius of the particle, and $\lambda_{\rm a}$ [cm] the atmospheric mean free path (Paper I).
The monodisperse Knudsen number, $\overline{{\rm Kn}}$, is simply given by
\begin{equation}
    \overline{{\rm Kn}} = \frac{\lambda_{\rm a}}{r_{\rm c}}.
\end{equation}
We can define a number-weighted population averaged Knudsen number, $\overline{{\rm Kn}_{N}}$, as
\begin{equation}
    \overline{{\rm Kn}_{N}}= \frac{\int_{0}^{\infty}{\rm Kn}(m)f(m)dm}{\int_{0}^{\infty}f(m)dm} = \frac{\lambda_{\rm a}b\int_{0}^{\infty}m^{-1/3}f(m)dm}{N_{\rm c}},
\end{equation}
where $b$ = $(3/4\pi\rho_{\rm d})^{-1/3}$ is the spherical geometric factor.
Using the definition of the moments and the Eq. \eqref{eq:mom_gen} moment generator, the number-weighted Knudsen number is
\begin{equation}
\label{eq:Kn_N}
    \overline{{\rm Kn}_{N}} = \frac{\lambda_{\rm a}bM^{(-1/3)}}{N_{\rm c}} = \frac{\lambda_{\rm a}}{r_{\rm c}}\nu^{1/3}\frac{\Gamma(\nu - 1/3)}{\Gamma(\nu)},
\end{equation}
where $M^{(-1/3)}$ is the $k$ = -1/3 non-integer moment power.
The $\nu^{1/3}$ factor in Eq. \eqref{eq:Kn_N} is given through using the relation in Eq. \eqref{eq:lambda} and combining $b$ and $m_{\rm c}$ to return the characteristic radius, $r_{\rm c}$, into the equation.
Similarly, a mass-weighted Knudsen number, $\overline{{\rm Kn}_{m}}$,  is given by
\begin{equation}
    \overline{{\rm Kn}_{m}} = \frac{\int_{0}^{\infty}{\rm Kn}(m)mf(m)dm}{\int_{0}^{\infty}mf(m)dm} = \frac{\lambda_{\rm a}b\int_{0}^{\infty}m^{2/3}f(m)dm}{\rho_{\rm c}},
\end{equation}
resulting in
\begin{equation}
    \overline{{\rm Kn}_{m}} = \frac{\lambda_{\rm a}bM^{(2/3)}}{\rho_{\rm c}} = \frac{\lambda_{\rm a}}{r_{\rm c}}\nu^{1/3}\frac{\Gamma(\nu + 2/3)}{\Gamma(\nu + 1)}.
\end{equation}
Finally, repeating the exercise for a weighting of $m^{2}$ gives
\begin{equation}
    \overline{{\rm Kn}_{m^{2}}} = \frac{\lambda_{\rm a}}{r_{\rm c}}\nu^{1/3}\frac{\Gamma(\nu + 5/3)}{\Gamma(\nu + 2)}.
\end{equation}

Figure \ref{fig:Kn_pop} shows how the population-averaged Knudsen numbers vary with $\nu$. 
Notably, the number-weighted formulation exhibits singularities for $\nu$ $\leq$ 1/3 due to the divergence of the gamma function.
To resolve this, we adopt an incomplete gamma formulation (Appendix \ref{app:corr}), which avoids the singularities and is implemented in the model code. 
The resulting solution is also shown in Figure \ref{fig:Kn_pop}.
As $\nu$ becomes larger and the distribution narrows, all three population averaged Knudsen numbers converge towards the monodisperse value.

In the continuum (diffusive) regime (Kn $\ll$ 1), the condensation or evaporation rate is given by \citep{Woitke2003}
\begin{equation}
\label{eq:cond_d}
    \left(\frac{dm}{dt}\right)^{\rm diff} = 4\pi r D m_{0} n_{\rm v} \left(1 - \frac{1}{S}\right),
\end{equation}
where $m_{0}$ [g] the mass of one unit of condensate, $n_{\rm v}$ [cm$^{-3}$] the number density of the vapour, $S$ the supersaturation ratio of the vapour and $D$ [cm$^{2}$ s$^{-1}$] the diffusive rate of the vapour onto the particle surface (Paper I).

The particle radius is converted to particle mass through the relation
\begin{equation}
    r = \left(\frac{3m}{4\pi\rho_{d}}\right)^{1/3},
\end{equation}
which reveals the mass dependence, $\propto$ $m^{1/3}$, of condensational growth and evaporation in the diffusive regime.
Defining the mass independent kinetic pre-factor as $C_{0}$
\begin{equation}
    C_{0} = 4\pi \left(\frac{3}{4\pi\rho_{d}}\right)^{1/3} D m_{0} n_{\rm v} \left(1 - \frac{1}{S}\right),
\end{equation}
equation Eq. \eqref{eq:dMk_dt} becomes
\begin{equation}
    \begin{split}
      \frac{dM^{(k)}}{dt} & =  kC_{0}\int_{0}^{\infty}m^{k-2/3}f(m)dm \\ & =  kC_{0}M^{(k - 2/3)}.
    \end{split}
\end{equation}
We can then substitute in the moment generator for the gamma distribution (Eq. \ref{eq:mom_gen})
\begin{equation}
    \frac{dM^{(k)}}{dt} = kC_{0}\lambda^{k-2/3}N_{\rm c}\frac{\Gamma(\nu + k - 2/3)}{\Gamma(\nu)}.
\end{equation}
We can then use Eq. \eqref{eq:mom_gen} again for the $k-1$ moment to factor out $N_{\rm c}$ and get the expression in terms of the previous moment
\begin{equation}
    \frac{dM^{(k)}}{dt} = kC_{0}\lambda^{1/3}M^{(k-1)}\frac{\Gamma(\nu + k - 2/3)}{\Gamma(\nu + k - 1)}.
\end{equation}
We can then use the relation $\lambda^{1/3}$ = m$_{\rm c}$$^{1/3}$$\nu$$^{-1/3}$ to reintroduce the representative particle size, $r_{\rm c}$, into the equation.
The full expression for the first moment ($k$ = 1) is then
\begin{equation}
\label{eq:qcdt_l_1}
    \left(\frac{d\rho_{\rm c}}{dt}\right)_{\rm cond}^{\rm diff} = 4\pi r_{\rm c} D m_{0} n_{\rm v} \left(1 - \frac{1}{S}\right)\nu^{-1/3}\frac{\Gamma(\nu + 1/3)}{\Gamma(\nu)}N_{\rm c},
\end{equation}
and the second moment ($k$ = 2) as
\begin{equation}
\label{eq:qcdt_l_2}
    \left(\frac{dZ_{\rm c}}{dt}\right)_{\rm cond}^{\rm diff} = 8\pi r_{\rm c} D m_{0} n_{\rm v} \left(1 - \frac{1}{S}\right)\nu^{-1/3}\frac{\Gamma(\nu + 4/3)}{\Gamma(\nu + 1)}\rho_{\rm c}.
\end{equation}

In the free molecular (or kinetic) regime (Kn $\gg$ 1), the mass dependence is $\propto$ $m^{2/3}$ as the condensation rate expression is given by \citep[e.g.][]{Woitke2003}
\begin{equation}
\label{eq:cond_f}
    \left(\frac{dm}{dt}\right)^{\rm free} = 4\pi r^{2}v_{\rm th}m_{0} n_{\rm v}\alpha \left(1 - \frac{1}{S}\right),
\end{equation}
where $v_{\rm th}$ = $\sqrt{k_{\rm b}T/2\pi m_{v}}$ [cm s$^{-1}$] is the thermal velocity of the condensable vapour, $m_{v}$ [g] the mass of the condensable vapour, and $\alpha$ = [0,1] the `sticking efficiency' of the vapour to the particle surface.
Again, defining the constant kinetic pre-factor $C_{0}$ as
\begin{equation}
    C_{0} = 4\pi \left(\frac{3}{4\pi\rho_{d}}\right)^{2/3}v_{\rm th}m_{0} n_{\rm v} \alpha \left(1 - \frac{1}{S}\right),
\end{equation}
the moment generator is then
\begin{equation}
    \begin{split}
      \frac{dM^{(k)}}{dt} & =  kC_{0}\int_{0}^{\infty}m^{k-1/3}f(m)dm \\ & =  kC_{0}M^{(k - 1/3)}.
    \end{split}
\end{equation}
Similar to the Kn $\ll$ 1 regime solution, using Eq. \eqref{eq:mom_gen} the final expression is given as
\begin{equation}
    \frac{dM^{(k)}}{dt} = kC_{0}\lambda^{k-1/3}N_{\rm c}\frac{\Gamma(k + \nu - 1/3)}{\Gamma(\nu)},
\end{equation}
and in terms of the $k-1$ moment
\begin{equation}
    \frac{dM^{(k)}}{dt} = kC_{0}\lambda^{2/3}M^{(k-1)}\frac{\Gamma(k + \nu - 1/3)}{\Gamma(k + \nu - 1)}.
\end{equation}
As with the Kn $\ll$ 1 regime, using the relation $\lambda^{2/3}$ = m$_{\rm c}$$^{2/3}$$\nu$$^{-2/3}$ returns the representative particle size, $r_{\rm c}$, to the expression.
This then gives for the full first moment ($k$ = 1) equation
\begin{equation}
\label{eq:qcdt_h_1}
    \left(\frac{d\rho_{\rm c}}{dt}\right)_{\rm cond}^{\rm free} = 4\pi r_{\rm c}^{2} v_{\rm th}m_{0} n_{\rm v} \left(1 - \frac{1}{S}\right)\nu^{-2/3}\frac{\Gamma(\nu + 2/3)}{\Gamma(\nu)}N_{\rm c},
\end{equation}
and second moment ($k$ = 2)
\begin{equation}
\label{eq:qcdt_h_2}
    \left(\frac{dZ_{\rm c}}{dt}\right)_{\rm cond}^{\rm free} = 8\pi r_{\rm c}^{2} v_{\rm th}m_{0} n_{\rm v} \left(1 - \frac{1}{S}\right)\nu^{-2/3}\frac{\Gamma(\nu + 5/3)}{\Gamma(\nu + 1)}\rho_{\rm c}.
\end{equation}

We adopt the same tanh-based interpolation function, $f(Kn')$, from Paper I to smoothly transition between the Kn $\ll$ 1 and Kn $\gg$ 1 regimes
\begin{multline}
\label{eq:2nd_int}
    \left(\frac{d\rho_{\rm c}}{dt}\right)_{\rm cond} = f(\overline{{\rm Kn'_{\rm m}}})\left(\frac{d\rho_{\rm c}}{dt}\right)_{\rm cond}^{\rm diff}  \\ + \left[1 - f(\overline{{\rm Kn'_{\rm m}}})\right] \left(\frac{d\rho_{\rm c}}{dt}\right)_{\rm cond}^{\rm free},
\end{multline}
\begin{multline}
\label{eq:3nd_int}
    \left(\frac{dZ_{\rm c}}{dt}\right)_{\rm cond} = f(\overline{{\rm Kn'_{\rm m^{2}}}})\left(\frac{dZ_{\rm c}}{dt}\right)_{\rm cond}^{\rm diff}  \\ + \left[1 - f(\overline{{\rm Kn'_{\rm m^{2}}}})\right] \left(\frac{dZ_{\rm c}}{dt}\right)_{\rm cond}^{\rm free},
\end{multline}
where $\overline{{\rm Kn'_{\rm m}}}$ = $\overline{\rm Kn_{\rm m}}$/Kn$_{cr, m}$ and $\overline{{\rm Kn'_{\rm m^{2}}}}$ = $\overline{\rm Kn_{\rm m^{2}}}$/Kn$_{cr, m^{2}}$, with Kn$_{cr, m}$ and Kn$_{cr, m^{2}}$ the critical Knudsen numbers.
The critical Knudsen numbers are determined through flux balance between the Kn $\ll$ 1 and Kn $\gg$ 1 limits, following \citet{Woitke2003}
\begin{equation}
    {\rm Kn}_{cr, m} = \frac{\lambda_{\rm a}}{r_{\rm c}} \frac{\left(d\rho_{\rm c}/dt\right)_{\rm cond}^{\rm free}}{\left(d\rho_{\rm c}/dt\right)_{\rm cond}^{\rm diff}},
\end{equation}
and
\begin{equation}
    {\rm Kn}_{cr, m^{2}} =  \frac{\lambda_{\rm a}}{{r_{\rm c}}} \frac{\left(dZ_{\rm c}/dt\right)_{\rm cond}^{\rm free}}{\left(dZ_{\rm c}/dt\right)_{\rm cond}^{\rm diff}},
\end{equation}
for the mass-weighted and mass-squared-weighted critical Knudsen numbers respectively.

\section{Collisional growth with gamma distribution}
\label{sec:gamma_coll}

In this section, we extend the derivation of collisional growth rates from Paper I to include expressions for the second moment in both Knudsen number regimes.
\citet{Drake1972} derived the mass moment generating form of the Smoluchowski equation for collisional growth as
\begin{equation}
\label{eq:coll_mom}
\begin{split}
    \frac{dM^{(k)}}{dt} = &\frac{1}{2}\int_{0}^{\infty}\int_{0}^{\infty}K(m,m')f(m)f(m') \\ & \times \left[(m + m')^{k} - m^{k} - m'^{k}\right]dmdm',
\end{split}
\end{equation}
where $m$ and $m'$ represent different individual masses in the particle mass distribution and $K(m,m')$ [cm$^{3}$ s$^{-1}$] the kernel function.
The zeroth moment ($k$ = 0) rate is 
\begin{eqnarray}
\label{eq:N_coll}
    \nonumber
    \frac{dM^{(0)}}{dt} &=& -\frac{1}{2}\int_{0}^{\infty}\int_{0}^{\infty}K(m,m')f(m)f(m')dmdm'\\
    &=& -\frac{1}{2}\overline{K(m,m')}N_{\rm c}^2,
\end{eqnarray}
the first ($k$ = 1) moment
\begin{equation}
\label{eq:rho_coll}
    \frac{dM^{(1)}}{dt} = 0,
\end{equation}
which is an expression of mass conservation during collisional growth, and the second ($k$ = 2) moment
\begin{eqnarray}
\label{eq:Z_coll}
    \nonumber
    \frac{dM^{(2)}}{dt} &=& \int_{0}^{\infty}\int_{0}^{\infty}K(m,m')mm'f(m)f(m')dmdm'\\
    &=&\overline{K_{m}(m,m')}\rho_{\rm c}^2,
\end{eqnarray}
where we have defined the number-weighted population averaged collision kernel \citep[c.f. Appendix C of][]{Moran2023}, given as
\begin{equation}
 \label{eq:pop_av_k}
    \begin{split}
       \overline{K(m,m')} &= \frac{\int_{0}^{\infty}\int_{0}^{\infty}K(m,m')f(m)f(m')dmdm'}{\int_{0}^{\infty}\int_{0}^{\infty}f(m)f(m')dmdm'} \\
       & =  \frac{1}{N_{\rm c}^{2}}\int_{0}^{\infty}\int_{0}^{\infty}K(m,m')f(m)f(m')dmdm'.
    \end{split}
\end{equation}
In addition, we require the mass-weighted population averaged kernel for the second moment
\begin{equation}
 \label{eq:pop_av_k_mass}
    \begin{split}
       \overline{K_{m}(m,m')} &= \frac{\int_{0}^{\infty}\int_{0}^{\infty}K(m,m')mm'f(m)f(m')dmdm'}{\int_{0}^{\infty}\int_{0}^{\infty}mm'f(m)f(m')dmdm'} \\
       & =  \frac{1}{\rho_{\rm c}^{2}}\int_{0}^{\infty}\int_{0}^{\infty}K(m,m')mm'f(m)f(m')dmdm'.
    \end{split}
\end{equation}

\subsection{Particle Kn $\ll$ 1 regime}

For collisional growth driven by Brownian motion (coagulation) in the Kn $\ll$ 1 regime, the collisional kernel is given by \citep{Chandrasekhar1943}
\begin{equation}
\label{eq:Diff_k}
    K(r,r') = 4\pi\left[D(r)  + D(r')\right](r + r'),
\end{equation}
where $r$ and $r'$ represent different particle radii in the distribution and $D(r)$ is the particle diffusion factor given by \citep{Chandrasekhar1943}
\begin{equation}
    D(r) = \frac{k_{\rm b}T\beta}{6\pi\eta_{\rm a}r},
\end{equation}
where $\eta_{\rm a}$ [g cm$^{-1}$ s$^{-1}$] is the atmospheric dynamical viscosity (Paper I), and $\beta$, the Cunningham slip factor, generally depends on the radius (and therefore mass) through scaling with Knudsen number in the functional form
\begin{equation}
    \beta = 1 + {\rm Kn}[A + Be^{-C/{\rm Kn}}].
\end{equation}
As in Paper I, we use the parameters experimentally measured between Kn $\approx$ 0.5-83 from \citet{Kim2005} 
\begin{equation}
    \beta = 1 + {\rm Kn}[1.165 + 0.483e^{-0.997/{\rm Kn}}],
\end{equation}
which is approximated to within $\approx$10\% in the linear form
\begin{equation}
\label{eq:b_lin}
    \beta = 1 + A{\rm Kn},
\end{equation}
where $A$ = 1.639, calculated from an optimised least-squares fitting to the \citet{Kim2005} coefficients.
Converting Eq. \eqref{eq:Diff_k}  to mass units gives the mass dependence of the kernel
\begin{equation}
\label{eq:Kcoag_Brown}
    K(m,m') = K_{0}(\beta m^{-1/3} + \beta'm'^{-1/3})(m^{1/3} + m'^{1/3}),
\end{equation}
where $K_{0}$ is the constant kinetic pre-factor given by
\begin{equation}
    K_{0} = \frac{2k_{\rm b}T}{3\eta_{\rm a}}.
\end{equation}

\subsubsection{Gamma distribution solution}

We can now attempt a zeroth and second moment collisional rate solution assuming a gamma distribution.
Following \citet{Moran2023}, the number density weighted population averaged kernel in this regime is given by
\begin{multline}
\label{eq:K_ll1_1}
     \overline{K(m,m')}  =  \frac{K_{0}}{N_{\rm c}^{2}}\int_{0}^{\infty}\int_{0}^{\infty}(\beta m^{-1/3} + \beta'm'^{-1/3})(m^{1/3} + m'^{1/3}) \\ \cdot f(m)f(m')dmdm',    
\end{multline}
and the mass-weighted averaged kernel is
\begin{multline}
\label{eq:K_ll1_1_mass}
     \overline{K_{m}(m,m')} = \frac{K_{0}}{\rho_{\rm c}^{2}}\int_{0}^{\infty}\int_{0}^{\infty}(\beta m^{-1/3} + \beta'm'^{-1/3})(m^{1/3} + m'^{1/3}) \\ \cdot mm'f(m)f(m')dmdm'.  
\end{multline}

As in Paper I, we can take into account the size variation with $\beta$ inside the integral assuming its linear form (Eq. \ref{eq:b_lin}).
This leads to an addition moment dependent correction term for the population averaged kernels 
\begin{multline}
    \overline{K(m,m')} = \frac{2K_{0}}{N_{\rm c}^{2}}\Biggl[M^{(0)}M^{(0)} + M^{(1/3)}M^{(-1/3)} \\ + \lambda_{\rm a}A\left(\frac{3}{4\pi\rho_{\rm d}}\right)^{-1/3}\left(M^{(0)}M^{(-1/3)} + M^{(1/3)}M^{(-2/3)}\right)\Biggr],
\end{multline}
and
\begin{multline}
    \overline{K_{m}(m,m')} = \frac{2K_{0}}{\rho_{\rm c}^{2}}\Biggl[M^{(1)}M^{(1)} + M^{(4/3)}M^{(2/3)} \\ + \lambda_{\rm a}A\left(\frac{3}{4\pi\rho_{\rm d}}\right)^{-1/3}\left(M^{(1)}M^{(2/3)} + M^{(4/3)}M^{(1/3)}\right)\Biggr].
\end{multline}
The change in the number density and second moment are then
\begin{multline}
\label{eq:N_coag_g_2}
    \left(\frac{dN_{\rm c}}{dt}\right)_{\rm coag} \approx -\frac{2k_{\rm b}T}{3\eta_{\rm a}} N_{\rm c}^{2}\Biggl[1 + \frac{\Gamma(\nu + 1/3)\Gamma(\nu - 1/3)}{\Gamma(\nu)^{2}} \\ + \frac{\lambda_{\rm a}}{r_{\rm c}}A\nu^{1/3}\left(\frac{\Gamma(\nu - 1/3)}{\Gamma(\nu)} + \frac{\Gamma(\nu + 1/3)\Gamma(\nu - 2/3)}{\Gamma(\nu)^{2}}\right) \Biggr],
\end{multline}
and
\begin{multline}
\label{eq:Z_coag_g_2}
    \left(\frac{dZ_{\rm c}}{dt}\right)_{\rm coag} \approx \frac{4k_{\rm b}T}{3\eta_{\rm a}} \rho_{\rm c}^{2}\Biggl[1 + \frac{\Gamma(\nu + 4/3)\Gamma(\nu + 2/3)}{\Gamma(\nu + 1)^{2}} \\ + \frac{\lambda_{\rm a}}{r_{\rm c}}A\nu^{1/3}\left(\frac{\Gamma(\nu + 2/3)}{\Gamma(\nu + 1)} + \frac{\Gamma(\nu + 4/3)\Gamma(\nu + 1/3)}{\Gamma(\nu + 1)^{2}}\right) \Biggr],
\end{multline}
respectively.

\subsection{Particle Kn $\gg$ 1 regime}

For Brownian coagulation in the Kn $\gg$ 1 regime, the collisional kernel is given by \citep[e.g.][]{Jacobson2005}
\begin{equation}
    K(r,r') = \pi\left(r + r'\right)^{2}\sqrt{\left(v(m)^{2} + v(m')^{2}\right)},
\end{equation}
where $v$ is the thermal velocity of the particle given by
\begin{equation}
    v = \sqrt{\frac{8k_{\rm b}T}{\pi m}}.
\end{equation}
Converting the kernel to mass units gives
\begin{equation}
\label{eq:k_gg1}
    K(m,m') = K_{0}\left(m^{1/3} + m'^{1/3}\right)^{2}\left(m^{-1} + m'^{-1}\right)^{1/2},
\end{equation}
where $K_{0}$ is the kinetic pre-factor
\begin{equation}
    K_{0} = \left(\frac{3}{4\pi\rho_{\rm d}}\right)^{2/3}\sqrt{8\pi k_{\rm b}T}.
\end{equation}
As in Paper I, to make the integration analytically tractable, we approximate the square root term as separable
\begin{equation}
\label{eq:approx}
    \left(m^{-1} + m'^{-1}\right)^{1/2} \approx H\left(m^{-1/2} + m'^{-1/2}\right),
\end{equation}
where $H$ is a fitting factor.
In Paper I, this fitting factor was found to be $H$ $\approx$ 0.85 through minimising the relative error between a suitable range of mass pairs.
However, Figure \ref{fig:gamma_diff} shows that the monodisperse limit is not recovered, even for large values of $\nu$ when assuming $H$ $\approx$ 0.85, which suggests an inaccuracy when using this value.
Figure \ref{fig:gamma_diff} shows that the lower limit of $H$ = 1/$\sqrt{2}$ reproduces the monodisperse limit in-line with the other collisional growth processes.
We therefore adopt $H$ = 1/$\sqrt{2}$ as a physically motivated parameter rather than fitting it.
Despite this, it is likely the optimal value of $H$ varies between 1/$\sqrt{2}$ and 1 dependent on the exact distribution parameters and specific dominant mass range of the distribution.
This suggest that the collisional rate will be slightly underestimated for broad distributions in our framework.

\subsubsection{Gamma distribution solution}

We can follow the same method in \citet{Moran2023} to derive a population averaged kernel function for the gamma distribution.
Placing Eq. \eqref{eq:k_gg1} into Eq. \eqref{eq:pop_av_k} gives
\begin{multline}
  \overline{K(m,m')} = \frac{K_{0}H}{N_{\rm c}^{2}} \int_{0}^{\infty}\int_{0}^{\infty}\left(m^{1/3} + m'^{1/3}\right)^{2}\left(m^{-1/2} + m'^{-1/2}\right)\\ 
  \cdot f(m)f(m')dmdm',
\end{multline}
and Eq. \eqref{eq:k_gg1} into Eq. \eqref{eq:pop_av_k_mass} gives
\begin{multline}
  \overline{K_{m}(m,m')} = \frac{K_{0}H}{\rho_{\rm c}^{2}}\int_{0}^{\infty}\int_{0}^{\infty}\left(m^{1/3} + m'^{1/3}\right)^{2}\left(m^{-1/2} + m'^{-1/2}\right) \\
  \cdot mm'f(m)f(m')dmdm',
\end{multline}
for the number density and mass-weighted population averaged kernel respectively.
Integrating the equations results in the averaged kernel given in terms of moment powers
\begin{multline}
    \overline{K(m,m')}  = \frac{2K_{0}H}{N_{\rm c}^{2}} \\ 
    \cdot\left(M^{(2/3)}M^{(-1/2)} + 2M^{(1/3)}M^{(-1/6)} + M^{(1/6)}M^{(0)}\right),
\end{multline}
where $M^{(2/3)}$, $M^{(-1/2)}$, $M^{(1/3)}$, $M^{(-1/6)}$ and $M^{(-1/6)}$ are the $k$ = 2/3, $k$ = -1/2, $k$ = 1/3, $k$ = -1/6 and $k$ = 1/6 non-integer powers of the moments respectively and
\begin{multline}
    \overline{K_{m}(m,m')} = \frac{2K_{0}H}{\rho_{\rm c}^{2}} \\ 
    \cdot\left(M^{(5/3)}M^{(1/2)} + 2M^{(4/3)}M^{(5/6)} + M^{(7/6)}M^{(1)}\right),
\end{multline}
where $M^{(5/3)}$, $M^{(1/2)}$, $M^{(4/3)}$, $M^{(5/6)}$ and $M^{(7/6)}$ are the $k$ = 5/3, $k$ = 1/2, $k$ = 4/3, $k$ = 5/6 and $k$ = 7/6 non-integer powers of the moments respectively.
Using the moment generator for the gamma mass distribution (Eq. \ref{eq:mom_gen}), results in
\begin{multline}
     \overline{K(m,m')} = 2K_{0}H\lambda^{1/6} \Biggl[\frac{\Gamma(\nu + 2/3)\Gamma(\nu - 1/2)}{\Gamma(\nu)^{2}} \\ 
     + \frac{2\Gamma(\nu + 1/3)\Gamma(\nu - 1/6)}{\Gamma(\nu)^{2}} + \frac{\Gamma(\nu + 1/6)}{\Gamma(\nu)}\Biggr],
\end{multline}
and 
\begin{multline}
     \overline{K_{m}(m,m')} =  2K_{0}H\lambda^{1/6} \Biggl[\frac{\Gamma(\nu + 5/3)\Gamma(\nu + 1/2)}{\Gamma(\nu + 1)^{2}} \\ + \frac{2\Gamma(\nu + 4/3)\Gamma(\nu + 5/6)}{\Gamma(\nu + 1)^{2}} + \frac{\Gamma(\nu + 7/6)}{\Gamma(\nu + 1)}\Biggr].
\end{multline}
Using the relation $\lambda^{1/6}$ = m$_{\rm c}$$^{1/6}$$\nu$$^{-1/6}$ = $m_{\rm c}^{2/3}$$m_{\rm c}^{-1/2}$$\nu$$^{-1/6}$, the representative particle size, $r_{\rm c}$, is returned, and the final expression for the change in number density is
\begin{multline}
\label{eq:N_coag_gam_2}
    \left(\frac{dN_{\rm c}}{dt}\right)_{\rm coag} \approx -H\sqrt{\frac{8\pi k_{\rm b} T}{m_{\rm c}}}r_{\rm c}^{2}N_{\rm c}^{2} \nu^{-1/6} \Biggl[\frac{\Gamma(\nu + 2/3)\Gamma(\nu - 1/2)}{\Gamma(\nu)^{2}} \\ +  \frac{2\Gamma(\nu + 1/3)\Gamma(\nu - 1/6)}{\Gamma(\nu)^{2}} + \frac{\Gamma(\nu + 1/6)}{\Gamma(\nu)}\Biggr],
\end{multline}
and for the second moment
\begin{multline}
\label{eq:Z_coag_gam_2}
    \left(\frac{dZ_{\rm c}}{dt}\right)_{\rm coag} \approx  2H\sqrt{\frac{8\pi k_{\rm b} T}{m_{\rm c}}}r_{\rm c}^{2}\rho_{\rm c}^{2} \nu^{-1/6} \Biggl[\frac{\Gamma(\nu + 5/3)\Gamma(\nu + 1/2)}{\Gamma(\nu+1)^{2}} \\ + \frac{2\Gamma(\nu + 4/3)\Gamma(\nu + 5/6)}{\Gamma(\nu+1)^{2}} + \frac{\Gamma(\nu + 7/6)}{\Gamma(\nu+1)}\Biggr].
\end{multline}

In Eqs. \eqref{eq:N_coag_g_2} and \eqref{eq:N_coag_gam_2}, negative arguments to the gamma function are possible when $\nu$ is accompanied with a negative fractional scaling.
This leads to singularities in the solution, similar to the number-weighted Knudsen number in Eq. \eqref{eq:Kn_N} (Fig. \ref{fig:Kn_pop}).
In Appendix \ref{app:corr}, we provide a physical explanation for this diverging behaviour as well as a derivation utilising the incomplete gamma function which avoids the singularities when using the gamma function.
This incomplete gamma function solution is ultimately used in the numerical model.
We use the diffusive Knudsen number interpolation function from Paper I \citep{Moran2022} to calculate coagulation rates in the intermediate Knudsen number regime (Kn $\sim$ 1) for the zeroth and second moments population averaged kernels.

\subsection{Gravitational coalescence}

For gravitational coalescence, the collisional growth kernel is given by \citep[e.g.][]{Jacobson2005}
\begin{equation}
   \label{eq:grav_kern}
    K(r,r') = \pi(r + r')^{2}|v_{\rm f}(r) - v_{\rm f}(r')|E,
\end{equation}
where $v_{\rm f}$ [cm s$^{-1}$] is the settling velocity of the particle.
Defining $\Delta v_{\rm f} = |v_{\rm f}(r) - v_{\rm f}(r')|$ we follow \citet{Ohno2017} and evaluate $\Delta v_{\rm f}$ with a parameter, $\epsilon$, that estimates the relative velocity of the particles from the representative particle size settling velocity, $v_{\rm f}(r_{\rm c})$, giving $\Delta v_{\rm f}$ $\approx$ $\epsilon$$v_{\rm f}(r_{\rm c})$. 
This is taken as $\epsilon$ $\approx$ 0.5 following the results of \citet{Sato2016}, who found this value to best reproduce the results of a collisional bin resolving model for grain growth in protoplanetary disks.

For a size distribution, $\overline{E}$ is the population averaged collisional efficiency factor, dependent on the population averaged Stokes number, $\overline{{\rm Stk}}$, 
\begin{equation}
    \overline{{\rm Stk}} = \frac{\Delta v_{\rm f} v_{\rm f}}{g\langle r\rangle} = \frac{\epsilon v_{\rm f}(\langle r\rangle)^{2}}{g\langle r\rangle},
\end{equation}
where $\langle r\rangle$ [cm] is the number averaged particle radius (Eq. \ref{eq:r_av}).
$\overline{E}$ is then given by \citep{Guillot2014}
\begin{equation}
     \overline{E} =
    \begin{cases}
      {\rm max}\left[0, 1 - 0.42\overline{{\rm Stk}}^{-0.75}\right]. & \overline{{\rm Kn}_{N}} < 1 \\
      1. &  \overline{{\rm Kn}_{N}} \geq 1 
    \end{cases}
\end{equation}
The population averages in the above equations are used for the zeroth moment rate of change.
For the second moment, the representative radius, $r_{\rm c}$, is used for the Stokes number and mass-weighted Knudsen number, $\overline{{\rm Kn}_{m}}$, for the delineation of $\overline{E}$.
The population averaged settling velocities using gamma distributions can be calculated following Appendix \ref{app:vf}.

\subsubsection{Gamma distribution}

For a gamma distribution, we make the assumption that $\Delta$$v_{\rm f}$ and $E$ are well defined at their respective representative particle size values for the zeroth and second moment, allowing them to be taken outside the integral.
However, we note this assumption may induce significant inaccuracy in the solution since $\Delta$$v_{\rm f}$ and $E$ vary by a large amount dependent on the individual particle mass pairs, especially for broad distributions.
Defining the kinetic pre-factor as 
\begin{equation}
    K_{0} = \left(\frac{3}{4\pi\rho_{\rm d}}\right)^{2/3}\pi\Delta v_{\rm f} E,
\end{equation}
the population averaged kernel is expressed as
\begin{multline}
  \overline{K(m,m')} = \frac{K_{0}}{N_{\rm c}^{2}} \int_{0}^{\infty}\int_{0}^{\infty}\left(m^{1/3} + m'^{1/3}\right)^{2} \\ 
  \cdot f(m)f(m')dmdm'.
\end{multline}
Solving the integral results in
\begin{equation}
      \overline{K(m,m')} = \frac{2K_{0}}{N_{\rm c}^{2}}\left(M^{(2/3)}M^{(0)} + M^{(1/3)}M^{(1/3)}\right),
\end{equation}
and
\begin{equation}
      \overline{K_{m}(m,m')} = \frac{2K_{0}}{\rho_{\rm c}^{2}}\left(M^{(5/3)}M^{(1)} + M^{(4/3)}M^{(4/3)}\right),
\end{equation}
for the zeroth and second moment population averaged kernels respectively.

Using the moment generator (Eq. \ref{eq:mom_gen}) we get for the number density averaged kernel
\begin{equation}
  \overline{K(m,m')} = 2K_{0}\lambda^{2/3}\left[\frac{\Gamma(\nu + 2/3)}{\Gamma(\nu)} + \frac{\Gamma(\nu + 1/3)^{2}}{\Gamma(\nu)^{2}}\right],
\end{equation}
and for the mass-weighted kernel
\begin{equation}
  \overline{K_{m}(m,m')} = 2K_{0}\lambda^{2/3}\left[\frac{\Gamma(\nu + 5/3)}{\Gamma(\nu+1)} + \frac{\Gamma(\nu + 4/3)^{2}}{\Gamma(\nu+1)^{2}}\right].
\end{equation}
Using $\lambda^{2/3}$ = $m_{\rm c}^{2/3}$$\nu^{-2/3}$, the representative particle size, $r_{\rm c}$, can be returned to the final expression.
The equation for the zeroth moment is then
\begin{multline}
    \left(\frac{dN_{\rm c}}{dt}\right)_{\rm coal} \approx -\pi r_{\rm c}^{2}\Delta v_{\rm f}  \overline{E} N_{\rm c}^{2} \\ \nu^{-2/3}\left[\frac{\Gamma(\nu + 2/3)}{\Gamma(\nu)} + \frac{\Gamma(\nu + 1/3)^{2}}{\Gamma(\nu)^{2}}\right],
\end{multline}
and for the second moment is
\begin{multline}
    \left(\frac{dZ_{\rm c}}{dt}\right)_{\rm coal} \approx 2\pi r_{\rm c}^{2}\Delta v_{\rm f}  \overline{E} \rho_{\rm c}^{2} \\ \nu^{-2/3}\left[\frac{\Gamma(\nu + 5/3)}{\Gamma(\nu+1)} + \frac{\Gamma(\nu + 4/3)^{2}}{\Gamma(\nu+1)^{2}}\right].
\end{multline}

\subsection{Summary of condensation and collisional growth results}

\begin{figure*}
    \centering
    \includegraphics[width=0.49\linewidth]{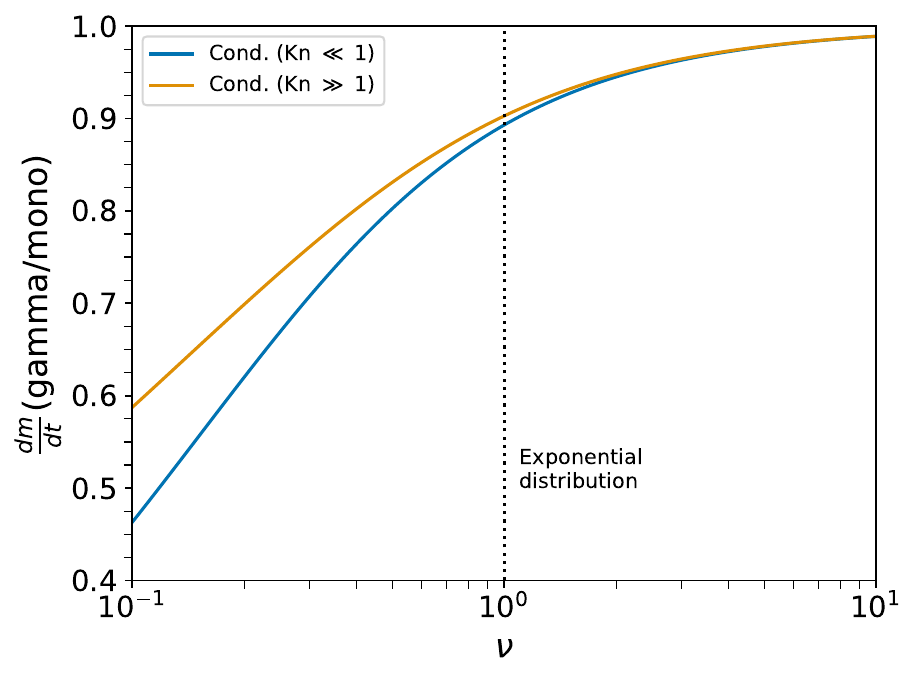}
    \includegraphics[width=0.49\linewidth]{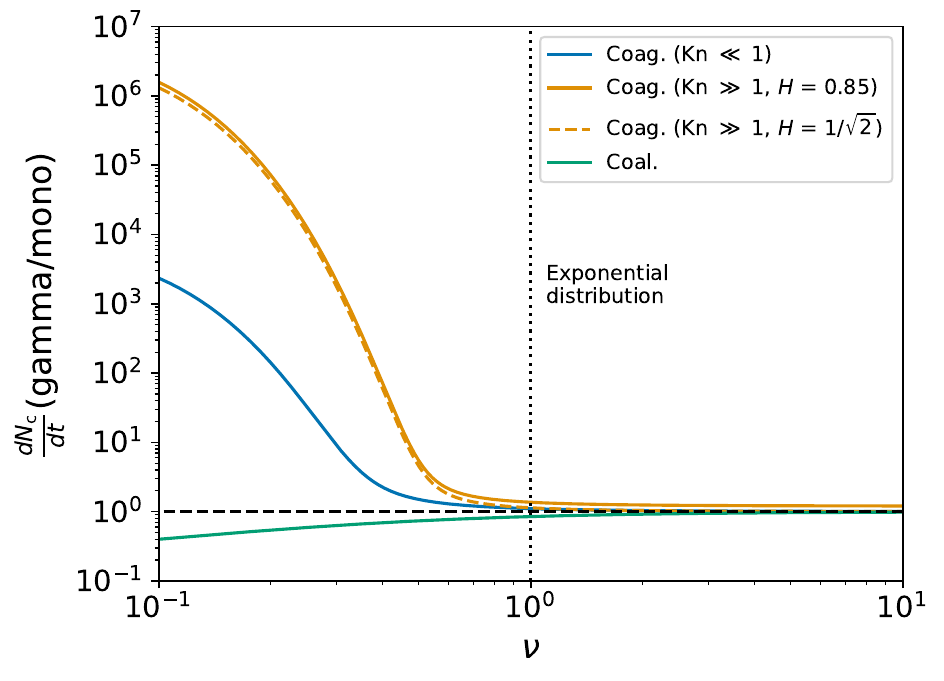}
    \caption{Relative factors between the gamma distribution and monodisperse distribution for the condensation and evaporation rates (left) and collisional growth rates (right).
    The vertical dotted line shows the $\nu$ = 1 values, denoting an exponential distribution (Paper I).
    As $\nu$ $\rightarrow$ $\infty$, the rates tend towards the monodisperse limit.
    However, for coagulation in the Kn $\gg$ 1 regime, assuming $H$ = 0.85 does not recover the monodisperse limit, even for large $\nu$, while a value of $H$ = 1/$\sqrt{2}$ does recover the monodisperse limit
}
    \label{fig:gamma_diff}
\end{figure*}

In Figure \ref{fig:gamma_diff}, we show the fractional difference between our derived gamma distribution expressions and the monodisperse values from \citet{Lee2025}.
For the condensation and evaporation rates, we find that the rates are reduced in the gamma distribution formulation by around a factor of two when $\nu$ = 0.1 compared to the monodisperse rates.
The rate converges towards the monodisperse solution when $\nu$ $\rightarrow$ $\infty$.

Figure \ref{fig:gamma_diff} also shows the fractional change in the coagulation and coalescence rates using the gamma distribution compared to the monodisperse formulation in \citet{Lee2025}.
Again, we find the rates converge towards the monodisperse solution when $\nu$ $\rightarrow$ $\infty$.
Generally, coalescence rates are reduced for broader distributions compared to the monodisperse case, and the coagulation rate is enhanced, sometimes by several magnitudes compared to the monodisperse case when the distribution is broad.
As stated earlier, for the coagulation rate, this plot shows that assuming the value of $H$ $\approx$ 0.85 does not recover the monodisperse limit, but the value of $H$ $\approx$ 1/$\sqrt{2}$ does. 
We therefore prefer a value of $H$ $\approx$ 1/$\sqrt{2}$ to ensure the monodisperse limit is recovered.
More rigorously, the optimal value of $H$ should depend on the dominant range of masses and should be a function of $\nu$ and $\lambda$, ranging between 1/$\sqrt{2}$ and 1.

For completeness, in Appendix \ref{sec:app_1}, we also derive population averaged kernels for the turbulence induced shear collisional rate.
We do not include the effects of turbulence induced collisions in the current study due to uncertainty in the atmospheric turbulent kinetic energy dissipation rate in sub-stellar atmospheres.
However, we note that calculations performed in \citet{Samra2022} suggest turbulence driven collisions may play an important role in setting the overall cloud structure in brown dwarf atmospheres.

\section{Y-dwarf KCl cloud simple 1D example}
\label{sec:1D}

\begin{figure*}
    \centering
    \includegraphics[width=0.48\linewidth]{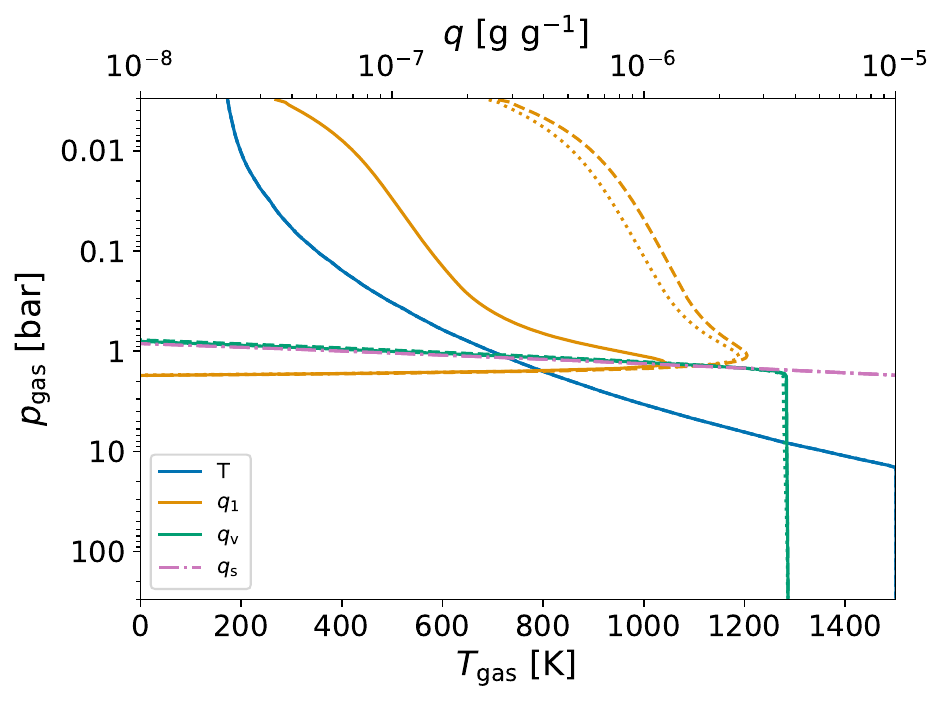}
    \includegraphics[width=0.48\linewidth]{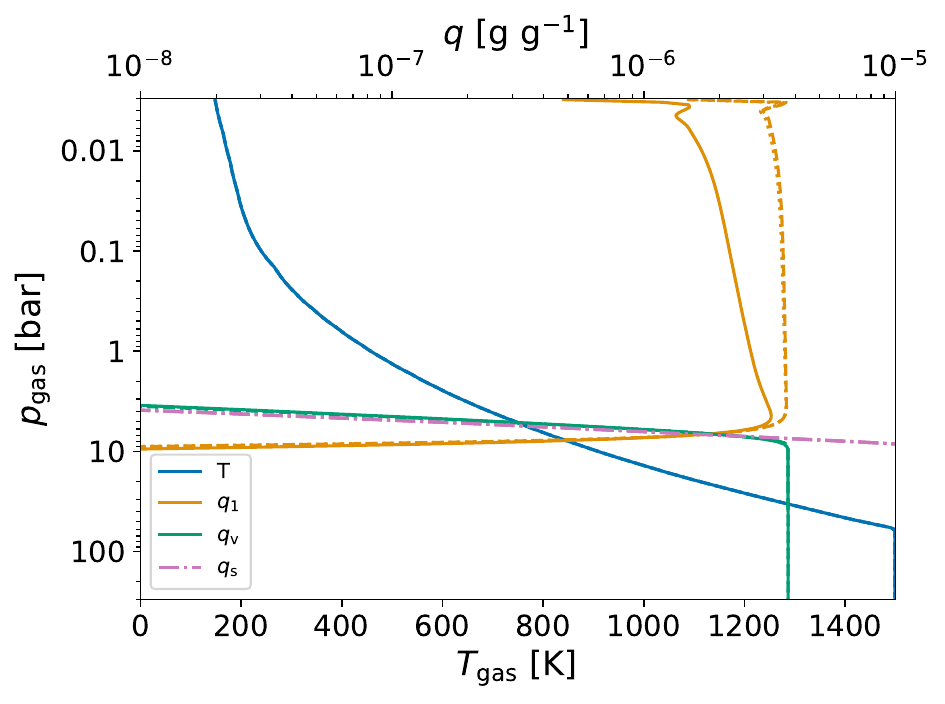}
    \includegraphics[width=0.48\linewidth]{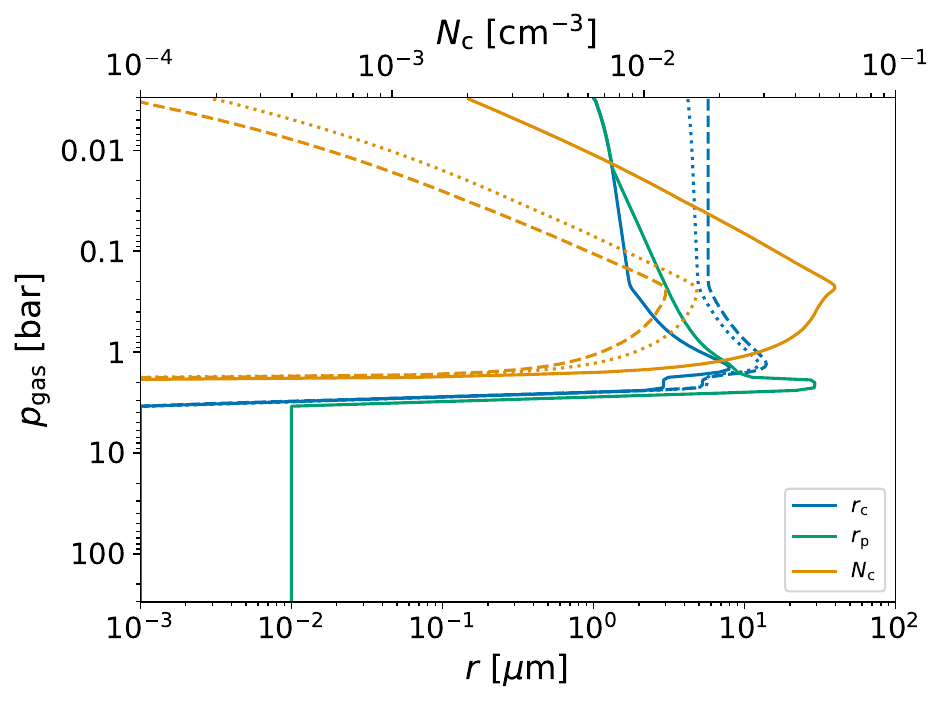}
    \includegraphics[width=0.48\linewidth]{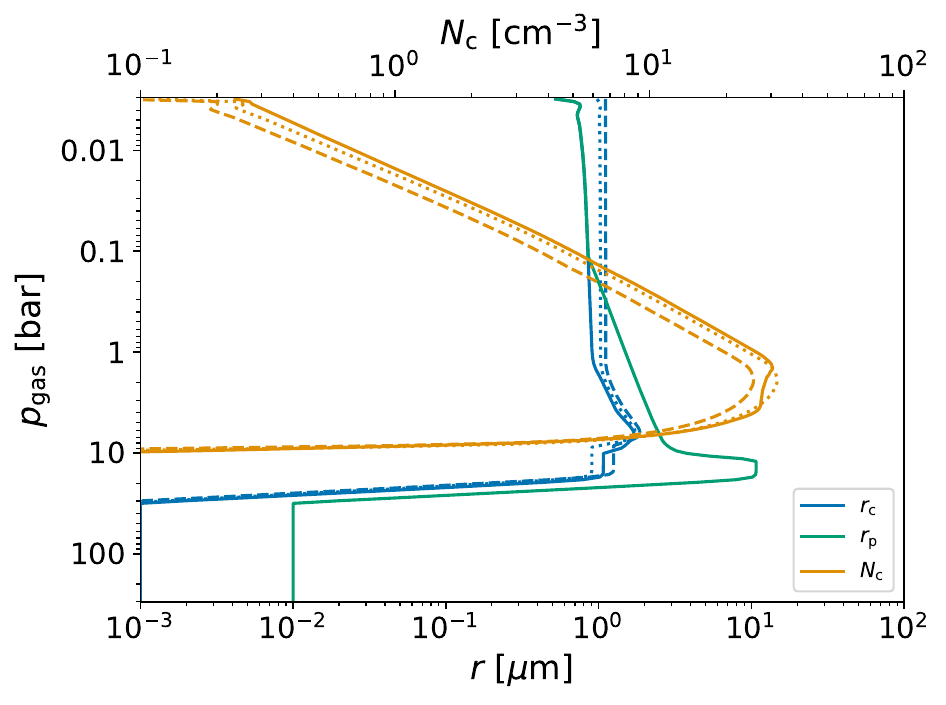}
    \includegraphics[width=0.48\linewidth]{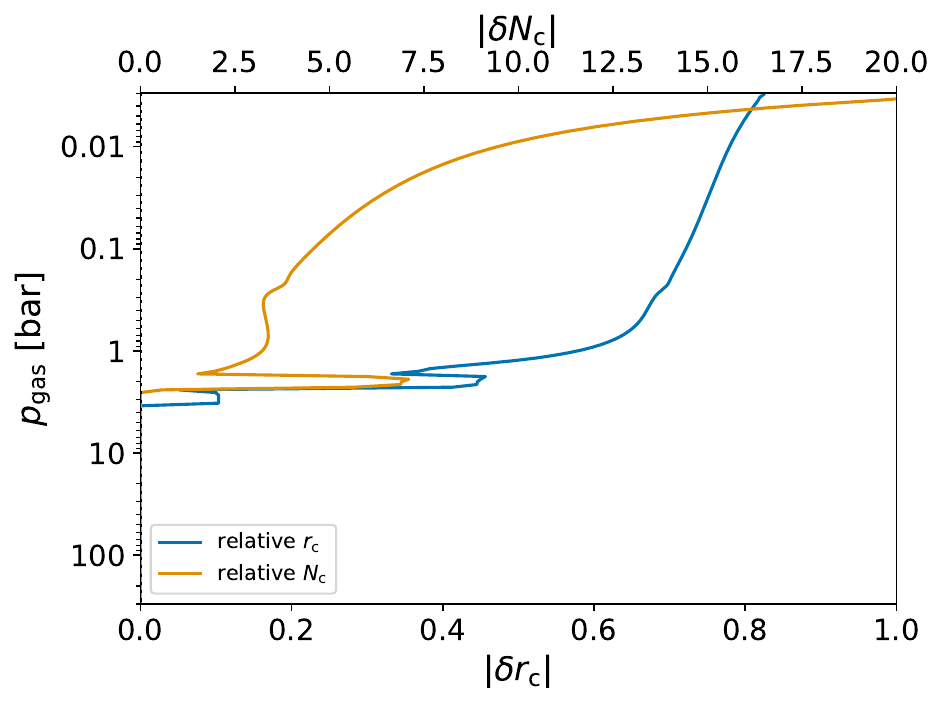}
    \includegraphics[width=0.48\linewidth]{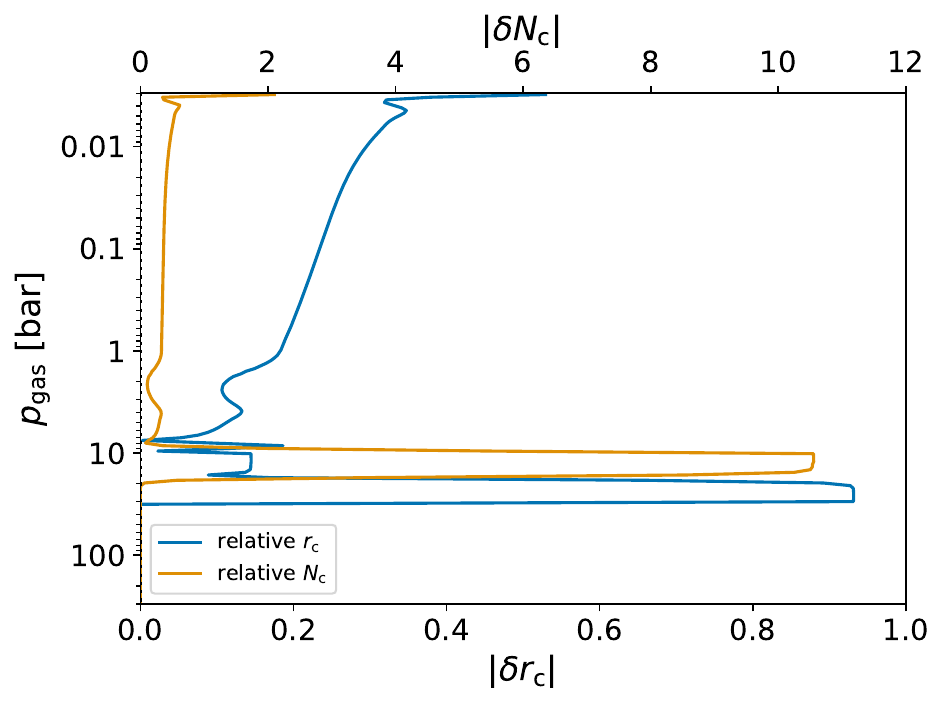}
    \caption{KCl cloud structures using the log g = 3.25 (left) and log g =  4.25 (right) Y-dwarf temperature-pressure profiles from \citet{Gao2018} and assuming a constant $K_{\rm zz}$ = 10$^{8}$ cm$^{2}$ s$^{-1}$.
    The dashed lines show the monodisperse size distribution, the dotted lines the exponential distribution, and the solid lines the exponential size distribution results.
    The top panel shows the temperature-pressure profiles with the mass mixing ratio of the condensate, $q_{\rm 1}$, vapour, $q_{\rm v}$ and saturation point, $q_{\rm s}$ (pink dash-dot line).
    The middle panel shows the representative particle sizes, $r_{\rm c}$ and $r_{\rm p}$, and total number density, $N_{\rm c}$. 
    The bottom panel shows the relative difference in $r_{\rm c}$ and $N_{\rm c}$ between the monodisperse and gamma cloud structure.}
    \label{fig:1D_KCl}
\end{figure*}

\begin{figure*}
    \centering
    \includegraphics[width=0.49\linewidth]{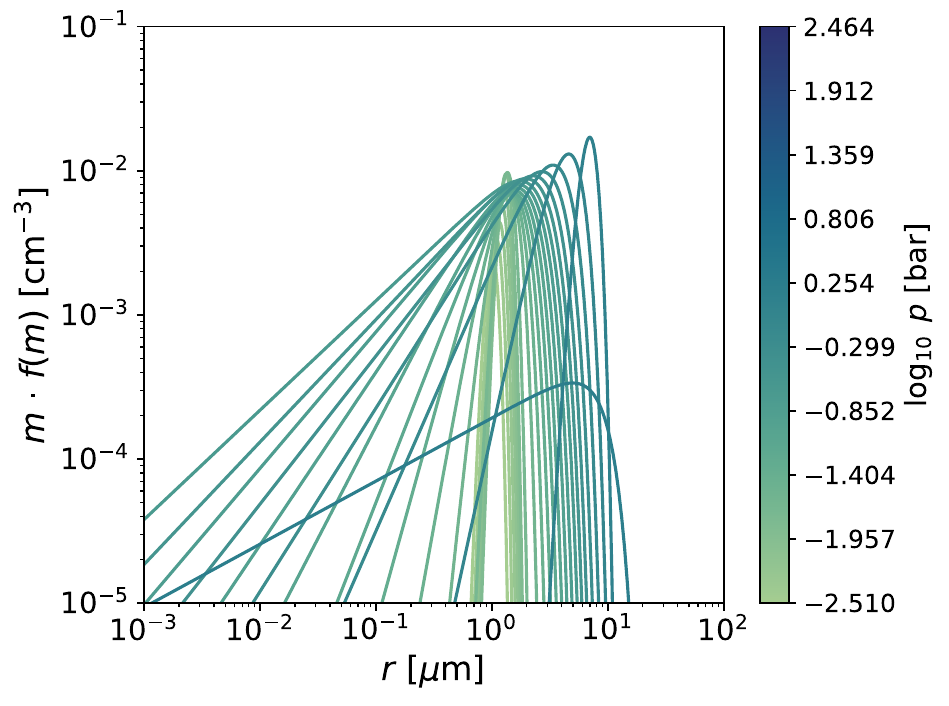}
    \includegraphics[width=0.49\linewidth]{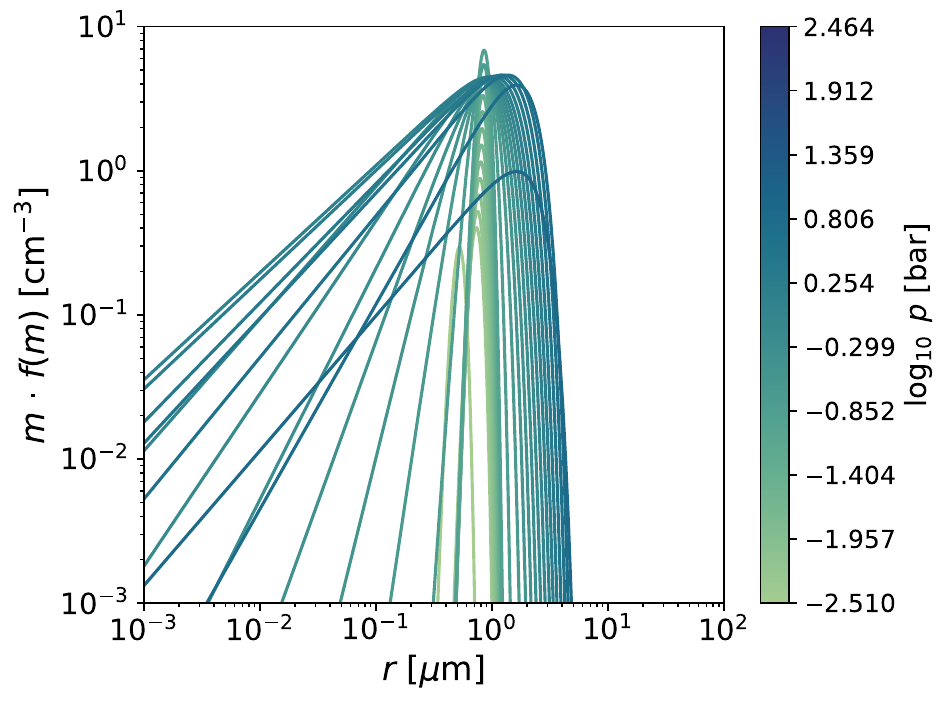}
    \caption{Reconstructed gamma particle size distributions from the simulations performed in Section \ref{sec:1D} for the log g = 3.25 (left) and log g = 4.25 (right) cases. 
    This shows the size distribution properties with pressure (colour bar) in the atmosphere. 
    In general, the size distribution narrows with decreasing pressure, with the most broad distributions occurring near the cloud base.}
    \label{fig:1D_dist}
\end{figure*}

In this section, we apply the three-moment gamma mass distribution framework to a simple 1D KCl Y-dwarf cloud formation example.
We use the same setup as Paper I, taking $T_{\rm eff}$ = 400 K, log g = 3.25 and log g = 4.25 Y-dwarf temperature-pressure (T-p) profiles from \citet{Gao2018} and assuming a constant $K_{\rm zz}$ = 10$^{8}$ cm$^{2}$ s$^{-1}$.
We include homogeneous nucleation of KCl cloud particles to form seed particles of size 1 nm following the same scheme as in \citet{Lee2025}.
In addition, we update the monodisperse and exponential distribution models from Paper I, where appropriate, using the methods above in order to create a one-to-one comparison as much as possible.
In particular, we now include moment dependent settling velocities following Appendix \ref{app:vf}, as well as a more accurate second order implicit Crank-Nicolson diffusion scheme and second order explicit TVD-MUSCL advection scheme.
We plot the absolute relative difference between the gamma and monodisperse results as
\begin{equation}
    |\delta N_{\rm c}| = \Biggl|\frac{N_{\rm c}({\rm gamma}) - N_{\rm c}({\rm mono})}{N_{\rm c}({\rm mono})}\Biggr|,
\end{equation}
with the equivalent expression for the representative particle size.

It is important to note that numerical issues appear in the formulation for large values of $\nu$ due to the inherent scalings present in the equations. 
First, we convert all equations outlined in this paper to numerically use the $\ln\Gamma(z)$ function rather than the $\Gamma(z)$ function in the code, which ensures numerical stability for larger $\nu$ values.
To achieve quick numerical convergence of the $\ln\Gamma(z)$ function, we assume a maximum of $\nu$ = 100 when the moment solutions give a highly monodisperse size distribution.
The size distribution is already shows highly monodisperse properties at $\nu$ = 10 (Fig. \ref{fig:gamma_diff}), suggesting this limit will have a negligible effect on the end results.

Noticeable differences are produced between the gamma and monodisperse results in some cases.
Figure \ref{fig:1D_KCl} shows our results for the 1D column model. 
In the log g = 3.25 case, the gamma distribution produces smaller particles by a factor of $\sim$2-3 and a larger number density by a factor of $\sim$10-20 compared to the monodisperse results.
This shows the potentially strong effects of polydispersity on the microphysical rates, as the distribution is quite broad across the deeper pressure levels, but narrows at lower pressure (Figure \ref{fig:1D_dist}).
In the log g = 4.25 case, fewer differences are seen between the gamma and monodisperse assumptions, with both $r_{\rm c}$ and $N_{\rm c}$ within around only a relative difference of only $\sim$0.3. 
The largest differences are seen at the cloud base, where the most polydisperse behavior is present, with the largest divergence between $r_{\rm c}$ and $r_{\rm p}$ occurring here.

We observe larger differences in the log g = 3.25 case, which is presumably because the gravitational separation of cloud particles is more effective.
As seen in Figure \ref{fig:1D_KCl}, the mass mixing ratio of clouds decreases with pressure in the log g = 3.25 case, indicating that some cloud particles have too fast settling velocity to be transported by eddy diffusion.
Meanwhile, we observe subtle differences ($<$20\%) among monodisperse, exponential, and gamma distribution models for the log g = 4.25 case, in which the gravitational separation barely occurs as informed by the vertically constant cloud abundance. 
Thus, our results suggest that polydipersity of the size distribution does matter when the gravitational separation of cloud particles play a role, as it controls the fraction of small particles that can be transported by eddy diffusion.
From Figure \ref{fig:1D_KCl}, the exponential distribution results are shifted towards the gamma distribution solution, compared to the monodisperse solution.
This suggests the exponential distribution succeeds in recovering some of the behaviour of the gamma distribution, although the inflexibility of the exponential distribution limits the full realisation of the microphysical effects from polydispersity.

In Figure \ref{fig:1D_dist}, we show the reconstructed gamma size distribution, derived from the moment solution values, as a function of pressure in the atmosphere for both test cases.
This shows how the size distribution is changing with altitude, with both cases showing a much broader size distribution near the cloud base which narrows to a more peaked distribution at lower pressure.
This suggests a differentiation of cloud particle sizes is occurring in the atmosphere, brought on by the relative differences in particle size settling velocity and microphysical effects.
Since the second moment settles faster than the first and zeroth moment, unless the distribution is very narrow, the variance of the distribution naturally reduces with pressure.
This allows the three-moment model to reproduce the trend of mean particle sizes decreasing with height for the log g = 3.25 case, which is seen in bin based microphysical models \citep[e.g.][]{Gao2018}, which is unable to be captured by the monodisperse model.

\section{Discussion}
\label{sec:disc}

Our results suggest that strong polydispersity is present at the cloud base, but as the cloud structure reaches lower pressures the cloud size distribution narrows, eventually showing highly monodisperse characteristics in the upper atmosphere.
We suggest this is a key benefit from the three-moment scheme, where the variance of the distribution is self-consistently calculated due to the addition of the second moment to the scheme.
In addition, we suggest the variation in settling velocity between the moments, where the second moment settles at the highest rate, is an important part of this processes, as reducing the value of $Z_{\rm c}$ relative to $\rho_{\rm c}$ and $N_{\rm c}$ decreases the variance of the distribution with height.
This is seen most strongly in Figure \ref{fig:1D_KCl} for the log g = 3.25 case, where a gradient towards smaller particles is produced in the representative particle size in the upper atmosphere for the gamma distribution, while the monodisperse results show a constant value.

\citet{Woitke2020} outline an approach to limit the coupling of cloud particles to the diffusive mixing of the atmosphere dependent on the local Stokes number of the particle relative to the largest turbulent eddy size.
In this study, we assumed all particles were equally affected by the vertical diffusive motions through the $K_{\rm zz}$ value.
We suggest that this Stokes number would also be moment dependent, resulting in further differentiation between the moment solutions, due to the first and second moments being more decoupled from the diffusive vertical mixing rate than the zeroth.
This would likely result in a further decrease in the variance and representative particle mass in the upper atmosphere compared to the results presented in this study.

Compared to Paper I, our gamma distribution results show better agreement with the microphysical bin-resolving model CARMA performed in \citet{Gao2018}.
For the log g = 3.25 case, the mixing ratio profile shows excellent agreement with the CARMA, as well as the representative particle size for the gamma distribution case now being smaller than the exponential and monodisperse case, more in line with the CARMA results.
However, for the log g = 4.25 case, CARMA recovers particle sizes in the $\sim$3 micron range, slightly larger than that produced here.
This may suggest strong multi-model behavior in this case, with the moment method capturing the smaller particle size population while CARMA contains a secondary larger particle sized mode.
We note that due to differences in the microphysical equations used, direct one-to-one comparison between our results and CARMA should be treated with caution. 
However, due to the similarly in overall cloud structures between the bin and bulk models, this suggests the bulk model is capturing well the complex behaviour of the bin model.
Despite this, we note that many bin models applied to exoplanet atmosphere clouds show strong evidence of bi- or multi-modal size distributions \citep[e.g.][]{Powell2019}, typically with a nucleation secondary component composed of small particles and a primary component of particles that have undergone condensational and collisional growth.
Possibly, to capture this effect using the moment method, multiple moments can be used to describe different parts of the distribution, such as the `cloud' and `rain' components used in \citet{Ohno2017}.
We note that such multi-moment schemes are standard practice for Earth water cloud microphysical models \citep[e.g.][]{Seifert2001}.

Our results from Figure \ref{fig:gamma_dist} suggest that the particle size distribution is likely to be well differentiated between the cloud base and cloud top, with a broad distribution at the cloud base and more peaked particle sizes at the cloud top.
We suggest that this behaviour fits well with current retrieval results for clouds in brown dwarf atmospheres \citep[e.g.][]{Burningham2021}, which best fit a broad particle size at the cloud base with a small particle component higher in the upper atmosphere.
With a broader particle size and increased number density, the cloud opacity is generally larger at the cloud base which decreases at the distribution becomes thinner and more monodisperse with a smaller particle size.
A broader distribution also reduces the strength of the infrared absorption features present in hot Jupiter and brown dwarf data \citep[e.g.][]{ Burningham2021, Grant2023}.
This is generally due to the contribution of larger particles in the distribution as shown in \citet{Wakeford2015}.
Should a more peaked size distribution be present in the observable upper atmosphere, emergent from a deeper atmosphere, broader distribution at the cloud base, this would enhance size specific cloud signatures at infrared wavelengths, as well as help produce a size-dependent scattering slope at optical wavelengths.

\section{Conclusions}
\label{sec:conc}

In this follow up study to \citet{Lee2025} and \citet{Lee2025p} (Paper I), we presented a self-consistent three-moment scheme for sub-stellar cloud microphysics including condensation, nucleation and collisional growth assuming a gamma size distribution.
Importantly, the present three-moment scheme is capable of simulating the evolution of the size distribution shape, which was fixed to the more inflexible monodisperse or exponential distributions throughout atmospheres in the two-moment scheme.
Overall, our derived equation set results in a modification to the monodisperse rates with the variance of the distribution also taken into account, going beyond the constant variance of the exponential distribution.
As a check, we show that our equation set trends towards the monodisperse solution when very narrow size distribution parameters are produced.
These modifications come in the form of simple additional gamma function terms and parameters that depend on the size distribution properties.
Here, we aimed to directly compare the monodisperse and polydisperse assumptions and isolate their effects on the cloud condensation, evaporation and collisional growth rates.

In our 1D KCl Y-dwarf cloud test, we found differences between the monodisperse and gamma distribution results, stemming from the effects of polydispersity altering condensation, Brownian coagulation and gravitational coalescence rates compared to the monodisperse rates.
Generally, Brownian coagulation rates are enhanced with polydispersity, while the gravitational coalescence rates are reduced.
We showed that the variance of the distribution changes with height, with the size distribution narrowing with altitude.
The log g = 3.25 case shows noticeable differences between the three-moment scheme and previous two-moment scheme with monodisperse and exponential distribution, suggesting that polydispersity plays a more important role for low gravity objects where the gravitational separation of cloud particles is more pronounced. 

With consideration of additional first moments for different condensation species, mixed material grains can be taken into account similar to those in \citet{Helling2008}, which is left to future studies.
Other three moment schemes utilising other size distribution assumptions, such as the log-normal distribution, can also be easily derived using our framework.

\begin{acknowledgements}
E.K.H. Lee is supported by the CSH through the Bernoulli Fellowship.
K. Ohno is supported by the JSPS KAKENHI Grant Number JP23K19072.
\end{acknowledgements}

\bibliographystyle{aa} 
\bibliography{bib.bib} 

\begin{thebibliography}{48}
\expandafter\ifx\csname natexlab\endcsname\relax\def\natexlab#1{#1}\fi

\bibitem[{{Ackerman} \& {Marley}(2001)}]{Ackerman2001}
{Ackerman}, A.~S. \& {Marley}, M.~S. 2001, \apj, 556, 872

\bibitem[{{Burningham} {et~al.}(2021){Burningham}, {Faherty}, {Gonzales}, {Marley}, {Visscher}, {Lupu}, {Gaarn}, {Fabienne Bieger}, {Freedman}, \& {Saumon}}]{Burningham2021}
{Burningham}, B., {Faherty}, J.~K., {Gonzales}, E.~C., {et~al.} 2021, \mnras, 506, 1944

\bibitem[{{Chandrasekhar}(1943)}]{Chandrasekhar1943}
{Chandrasekhar}, S. 1943, Reviews of Modern Physics, 15, 1

\bibitem[{{Christie} {et~al.}(2022){Christie}, {Mayne}, {Gillard}, {Manners}, {H{\'e}brard}, {Lines}, \& {Kohary}}]{Christie2022}
{Christie}, D.~A., {Mayne}, N.~J., {Gillard}, R.~M., {et~al.} 2022, \mnras, 517, 1407

\bibitem[{{Drake}(1972)}]{Drake1972}
{Drake}, R.~L. 1972, Journal of the Atmospheric Sciences, 29, 537

\bibitem[{Gail \& Sedlmayr(2013)}]{Gail2013}
Gail, H.-P. \& Sedlmayr, E. 2013, Physics and Chemistry of Circumstellar Dust Shells, Cambridge Astrophysics (Cambridge University Press)

\bibitem[{{Gao} {et~al.}(2018){Gao}, {Marley}, \& {Ackerman}}]{Gao2018}
{Gao}, P., {Marley}, M.~S., \& {Ackerman}, A.~S. 2018, \apj, 855, 86

\bibitem[{{Garrett}(2019)}]{Garrett19}
{Garrett}, T.~J. 2019, Journal of the Atmospheric Sciences, 76, 1031

\bibitem[{{Grant} {et~al.}(2023){Grant}, {Lewis}, {Wakeford}, {Batalha}, {Glidden}, {Goyal}, {Mullens}, {MacDonald}, {May}, {Seager}, {Stevenson}, {Valenti}, {Visscher}, {Alderson}, {Allen}, {Ca{\~n}as}, {Col{\'o}n}, {Clampin}, {Espinoza}, {Gressier}, {Huang}, {Lin}, {Long}, {Louie}, {Pe{\~n}a-Guerrero}, {Ranjan}, {Sotzen}, {Valentine}, {Anderson}, {Balmer}, {Bellini}, {Hoch}, {Kammerer}, {Libralato}, {Mountain}, {Perrin}, {Pueyo}, {Rickman}, {Rebollido}, {Sohn}, {van der Marel}, \& {Watkins}}]{Grant2023}
{Grant}, D., {Lewis}, N.~K., {Wakeford}, H.~R., {et~al.} 2023, \apjl, 956, L32

\bibitem[{{Guillot} {et~al.}(2014){Guillot}, {Ida}, \& {Ormel}}]{Guillot2014}
{Guillot}, T., {Ida}, S., \& {Ormel}, C.~W. 2014, \aap, 572, A72

\bibitem[{{Hansen} \& {Hovenier}(1974)}]{Hansen&Hovenier74}
{Hansen}, J.~E. \& {Hovenier}, J.~W. 1974, Journal of the Atmospheric Sciences, 31, 1137

\bibitem[{{Helling} {et~al.}(2008){Helling}, {Woitke}, \& {Thi}}]{Helling2008}
{Helling}, C., {Woitke}, P., \& {Thi}, W.~F. 2008, \aap, 485, 547

\bibitem[{{Hiranaka} {et~al.}(2016){Hiranaka}, {Cruz}, {Douglas}, {Marley}, \& {Baldassare}}]{Hiranaka+16}
{Hiranaka}, K., {Cruz}, K.~L., {Douglas}, S.~T., {Marley}, M.~S., \& {Baldassare}, V.~F. 2016, \apj, 830, 96

\bibitem[{Jacobson(2005)}]{Jacobson2005}
Jacobson, M.~Z. 2005, Fundamentals of Atmospheric Modeling, 2nd edn. (Cambridge University Press)

\bibitem[{Kim {et~al.}(2005)Kim, Mulholland, Kukuck, \& Pui}]{Kim2005}
Kim, J., Mulholland, G., Kukuck, S., \& Pui, D. 2005, Journal of Research of the National Institute of Standards and Technology, 110

\bibitem[{{Komacek} {et~al.}(2022){Komacek}, {Tan}, {Gao}, \& {Lee}}]{Komacek2022}
{Komacek}, T.~D., {Tan}, X., {Gao}, P., \& {Lee}, E. K.~H. 2022, \apj, 934, 79

\bibitem[{{Lee} {et~al.}(2016){Lee}, {Dobbs-Dixon}, {Helling}, {Bognar}, \& {Woitke}}]{Lee2016}
{Lee}, E., {Dobbs-Dixon}, I., {Helling}, C., {Bognar}, K., \& {Woitke}, P. 2016, \aap, 594, A48

\bibitem[{{Lee}(2023)}]{Lee2023}
{Lee}, E. K.~H. 2023, \mnras, 524, 2918

\bibitem[{{Lee}(2025)}]{Lee2025p}
{Lee}, E. K.~H. 2025, \aap, 698, A220

\bibitem[{{Lee} \& {Ohno}(2025)}]{Lee2025}
{Lee}, E. K.~H. \& {Ohno}, K. 2025, \aap, 695, A111

\bibitem[{{Lines} {et~al.}(2018){Lines}, {Mayne}, {Boutle}, {Manners}, {Lee}, {Helling}, {Drummond}, {Amundsen}, {Goyal}, {Acreman}, {Tremblin}, \& {Kerslake}}]{Lines2018}
{Lines}, S., {Mayne}, N.~J., {Boutle}, I.~A., {et~al.} 2018, \aap, 615, A97

\bibitem[{Marshall \& Palmer(1948)}]{Marshall1948}
Marshall, J.~S. \& Palmer, W. M.~K. 1948, Journal of Atmospheric Sciences, 5, 165

\bibitem[{McFarquhar {et~al.}(2015)McFarquhar, Hsieh, Freer, Mascio, \& Jewett}]{McFarquhar+15}
McFarquhar, G.~M., Hsieh, T.~L., Freer, M., Mascio, J., \& Jewett, B.~F. 2015, J. Atmos. Sci., 72, 892

\bibitem[{{Milbrandt} \& {Yau}(2005)}]{Milbrandt&Yau05_3moment}
{Milbrandt}, J.~A. \& {Yau}, M.~K. 2005, Journal of the Atmospheric Sciences, 62, 3065

\bibitem[{{Mor{\'a}n} \& {Kholghy}(2023)}]{Moran2023}
{Mor{\'a}n}, J. \& {Kholghy}, M.~R. 2023, Aerosol Science Technology, 57, 782

\bibitem[{Morrison \& Gettelman(2008)}]{Morrison2008}
Morrison, H. \& Gettelman, A. 2008, Journal of Climate, 21, 3642

\bibitem[{Morán(2022)}]{Moran2022}
Morán, J. 2022, Fractal and Fractional, 6

\bibitem[{Naumann \& Seifert(2016)}]{Naumann&Seifert16}
Naumann, A. \& Seifert, A. 2016, J. Atmos. Sci., 73, 2279

\bibitem[{{Ohno} \& {Okuzumi}(2017)}]{Ohno2017}
{Ohno}, K. \& {Okuzumi}, S. 2017, \apj, 835, 261

\bibitem[{{Ohno} \& {Okuzumi}(2018)}]{Ohno2018}
{Ohno}, K. \& {Okuzumi}, S. 2018, \apj, 859, 34

\bibitem[{{Ohno} {et~al.}(2020){Ohno}, {Okuzumi}, \& {Tazaki}}]{Ohno2020}
{Ohno}, K., {Okuzumi}, S., \& {Tazaki}, R. 2020, \apj, 891, 131

\bibitem[{{Parmentier} {et~al.}(2016){Parmentier}, {Fortney}, {Showman}, {Morley}, \& {Marley}}]{Parmentier2016}
{Parmentier}, V., {Fortney}, J.~J., {Showman}, A.~P., {Morley}, C., \& {Marley}, M.~S. 2016, \apj, 828, 22

\bibitem[{Paukert {et~al.}(2019)Paukert, Fan, Rasch, Morrison, Milbrandt, Shpund, \& Khain}]{Paukert+19}
Paukert, M., Fan, J., Rasch, P.~J., {et~al.} 2019, Journal of Advances in Modeling Earth Systems, 11, 257

\bibitem[{{Powell} {et~al.}(2019){Powell}, {Louden}, {Kreidberg}, {Zhang}, {Gao}, \& {Parmentier}}]{Powell2019}
{Powell}, D., {Louden}, T., {Kreidberg}, L., {et~al.} 2019, \apj, 887, 170

\bibitem[{{Powell} \& {Zhang}(2024)}]{Powell2024}
{Powell}, D. \& {Zhang}, X. 2024, \apj, 969, 5

\bibitem[{{Press} {et~al.}(1986){Press}, {Flannery}, \& {Teukolsky}}]{Press1986}
{Press}, W.~H., {Flannery}, B.~P., \& {Teukolsky}, S.~A. 1986, {Numerical recipes. The art of scientific computing} (Cambridge University Press)

\bibitem[{{Roman} \& {Rauscher}(2019)}]{Roman2019}
{Roman}, M. \& {Rauscher}, E. 2019, \apj, 872, 1

\bibitem[{{Saffman} \& {Turner}(1956)}]{Saffman1956}
{Saffman}, P.~G. \& {Turner}, J.~S. 1956, Journal of Fluid Mechanics, 1, 16

\bibitem[{{Samra} {et~al.}(2022){Samra}, {Helling}, \& {Birnstiel}}]{Samra2022}
{Samra}, D., {Helling}, C., \& {Birnstiel}, T. 2022, \aap, 663, A47

\bibitem[{{Samra} {et~al.}(2020){Samra}, {Helling}, \& {Min}}]{Samra+20}
{Samra}, D., {Helling}, C., \& {Min}, M. 2020, \aap, 639, A107

\bibitem[{{Sato} {et~al.}(2016){Sato}, {Okuzumi}, \& {Ida}}]{Sato2016}
{Sato}, T., {Okuzumi}, S., \& {Ida}, S. 2016, \aap, 589, A15

\bibitem[{{Seifert} \& {Beheng}(2001)}]{Seifert2001}
{Seifert}, A. \& {Beheng}, K.~D. 2001, Atmospheric Research, 59, 265

\bibitem[{Straka(2009)}]{Straka2009}
Straka, J.~M. 2009, Cloud and Precipitation Microphysics: Principles and Parameterizations (Cambridge University Press)

\bibitem[{{Tan} \& {Showman}(2019)}]{Tan2019}
{Tan}, X. \& {Showman}, A.~P. 2019, \apj, 874, 111

\bibitem[{{Ulbrich}(1983)}]{Ulbrich83}
{Ulbrich}, C.~W. 1983, Journal of Applied Meteorology, 22, 1764

\bibitem[{{Wakeford} \& {Sing}(2015)}]{Wakeford2015}
{Wakeford}, H.~R. \& {Sing}, D.~K. 2015, \aap, 573, A122

\bibitem[{{Woitke} \& {Helling}(2003)}]{Woitke2003}
{Woitke}, P. \& {Helling}, C. 2003, \aap, 399, 297

\bibitem[{{Woitke} {et~al.}(2020){Woitke}, {Helling}, \& {Gunn}}]{Woitke2020}
{Woitke}, P., {Helling}, C., \& {Gunn}, O. 2020, \aap, 634, A23

\end{thebibliography}

\appendix

\section{Cause of gamma function divergence issue}
\label{app:corr}

In Section \ref{sec:gamma_coll}, we show that the coagulation rate of Eqs. \eqref{eq:N_coag_g_2} and \eqref{eq:N_coag_gam_2} derived from the gamma distribution includes a gamma function that potentially involves a zero or negative argument.
This seems problematic, as the coagulation rate diverges to infinity at $\nu\rightarrow1/3$ and $\nu\rightarrow2/3$ in Eq. \eqref{eq:N_coag_g_2}, which is clearly unrealistic.
This odd behavior arises because we have ignored the fact that the particle size distribution actually has a cut off at a certain small particle size.
For example, cloud particles cannot be smaller than the size of constituting molecules. 
In practice, enhanced evaporation due to particle curvature (Kelvin effect) would prohibit the presence such small particles.

To make the problem clearer, let us consider a thought experiment: a huge single particle is embedded in swarm of cloud particles.
Then, the rate of decrease in total number density of cloud particles due to collision between a single particle with a mass of $M$ and cloud particles.
For the Brownian coagulation in the continuum regime (${\rm Kn}\gg1$), one can describe the rate as
\begin{eqnarray}
    \nonumber
    \frac{dN_{\rm c}}{dt}&=&-\int_{\rm m_{\rm min}}^{\rm m_{\rm max}}K(m,M)f(m)dm\\
    &=&-\frac{K_{\rm 0}N_{\rm c}M^{1/3}}{\lambda^{\nu}\Gamma(\nu)}\int_{\rm m_{\rm min}}^{\rm m_{\rm max}}m^{\nu-4/3}e^{-m/\lambda}dm,
\end{eqnarray}
where the particle size distribution is postulated to have maximum and minimum masses of $m_{\rm max}$ [g] and $m_{\rm min}$ [g], and we have used Eqs. \eqref{eq:exp_dist} and \eqref{eq:Kcoag_Brown} with $M{\gg}m_{\rm max}$.
We can see that the contribution from the collisions of particles with masses of $m$ is then 
\begin{equation}
\Delta N_{\rm c}(m)=m^{\nu-4/3}e^{-m/\lambda}\Delta m\sim m^{\nu-1/3}e^{-m/\lambda}.
\end{equation}
In this equation, the contribution from smallest particles is completely different between the cases of $\nu>1/3$ and $\nu<1/3$.
$\Delta N_{\rm c}$ decreases with decreasing particle mass at $\nu>1/3$, implying that smallest particles barely contribute to the total collision rate, which justifies $m_{\rm min}=0$ for $\nu>1/3$.
However, for $\nu<1/3$, $\Delta N_{\rm c}$ turns out to increase with decreasing particle mass, implying that the collisions of smallest particles with $m=m_{\rm min}$ have the largest contribution to the total collision rate.
In this case, one cannot set $m_{\rm min}=0$, as the collision rate does not converge to zero and diverges to infinity at smallest particle mass.
Physically, this behavior happens because collision velocity is faster at smaller particles, and the size distribution needs to drop faster than the rise of collision velocity to make the collision rate being zero at $m=0$.
Note that the exponential cutoff $e^{-m/\lambda}$ guarantees that the collision rate always converges to zero at large size limit, justifying to set $m_{\rm max}=\infty$ for any $\nu$.

\subsection{Prescription to correct gamma function negative arguments}

To fix the divergence issue, one can introduce a prescription that accounts for the finite value of the minimum particle mass $m_{\rm min}$.
Since divergence issues are present for the population-averaged kernel $\overline{K(m,m')}$ due to the reasons explained above, we need only to apply the prescription to terms in $\overline{K(m,m')}$ that diverge.
Supposing that the collisions of particles with $m<m_{\rm min}$ do not occur in reality, we can rewrite the population averaged collision kernel as
\begin{eqnarray}
       \overline{K(m,m')} &=&  \frac{1}{N_{\rm c}^{2}}\int_{m_{\rm min}}^{\infty}\int_{m_{\rm min}}^{\infty}K(m,m')f(m)f(m')dmdm',
\end{eqnarray}
For the later convenience, we define the function
\begin{equation}\label{eq:I_generator}
    I^{(k)}=\int_{\rm m_{\rm min}}^{\rm \infty}m^{k}f(m)dm=N_{\rm c}\lambda^{k}\frac{\Gamma(\nu+k,x_{\rm min})}{\Gamma(\nu)},
\end{equation}
where $x_{\rm min}=m_{\rm min}/\lambda$ is the dimensionless minimum mass, and $\Gamma(s,x)\equiv\int_{\rm x}^{\rm \infty}t^{s-1}e^{-t}dt$ is the upper incomplete gamma function.
This function returns the $k$-th moment from Eq. \eqref{eq:mom_gen} as $x_{\rm min}\rightarrow 0$.
In our current models, $x_{\rm min}$ is taken to be the mass of particles at the seed particle size of 1 nm.

Here, we repeat the derivation of the averaged kernel with the above prescription.
For the Brownian coagulation in the particle diffusive regime (Kn $\ll$ 1), the averaged kernel (Eqs. \ref{eq:K_ll1_1} and \ref{eq:K_ll1_1_mass}) can be rewritten as
\begin{multline}
     \overline{K(m,m')}  =  \frac{K_{0}}{N_{\rm c}^{2}}\int_{m_{\rm min}}^{\infty}\int_{m_{\rm min}}^{\infty}[\beta m^{-1/3} + \beta'm'^{-1/3}](m^{1/3} + m'^{1/3})\\\cdot f(m)f(m')dmdm'.
\end{multline}
Adopting the linear slip factor of $\beta=1+A{\rm Kn}$ (Eq.\ref{eq:b_lin}), the population-averaged kernel is given by
\begin{multline}
    \overline{K(m,m')} = \frac{2K_{0}}{N_{\rm c}^{2}}\Biggl[I^{(0)}I^{(0)} + I^{(1/3)}I^{(-1/3)} \\ + \lambda_{\rm a}A\left(\frac{3}{4\pi\rho_{\rm d}}\right)^{-1/3}\left(I^{(0)}I^{(-1/3)} + I^{(1/3)}I^{(-2/3)}\right)\Biggr].
\end{multline}
Applying the generator function of Eq. \eqref{eq:I_generator} and inserting the population-averaged kernel back to Eq. \eqref{eq:N_coll}, the rate of change in number density is given by
\begin{multline}
\label{eq:fix_1}
    \left(\frac{dN_{\rm c}}{dt}\right)_{\rm coag} \approx -\frac{2k_{\rm b}T}{3\eta_{\rm a}} N_{\rm c}^{2}\Biggl[\frac{\Gamma(\nu,x_{\rm min})^2}{\Gamma(\nu)^2} + \frac{\Gamma(\nu + 1/3,x_{\rm min})\Gamma(\nu - 1/3,x_{\rm min})}{\Gamma(\nu)^{2}} \\ + \frac{\lambda_{\rm a}}{r_{\rm c}}A\nu^{1/3}\Biggl(\frac{\Gamma(\nu,x_{\rm min})\Gamma(\nu - 1/3,x_{\rm min})}{\Gamma(\nu)^2}  \\ + \frac{\Gamma(\nu + 1/3,x_{\rm min})\Gamma(\nu - 2/3,x_{\rm min})}{\Gamma(\nu)^{2}}\Biggr) \Biggr].
\end{multline}

For the kinetic regime of Brownian motion, the population-averaged kernel can be rewritten as
\begin{multline}
  \overline{K(m,m')} = \frac{K_{0}H}{N_{\rm c}^{2}} \int_{m_{\rm min}}^{\infty}\int_{m_{\rm min}}^{\infty}\left(m^{1/3} + m'^{1/3}\right)^{2}\left(m^{-1/2} + m'^{-1/2}\right)\\ 
  \cdot f(m)f(m')dmdm'.
\end{multline}
This equation yields
\begin{multline}
    \overline{K(m,m')}  = \frac{2K_{0}H}{N_{\rm c}^{2}} \left(I^{(2/3)}I^{(-1/2)} + 2I^{(1/3)}I^{(-1/6)} + I^{(1/6)}I^{(0)}\right).
\end{multline}
Again, using the generator function of Eq. \eqref{eq:I_generator} and inserting the population-averaged kernel back to Eq. \eqref{eq:N_coll}, the rate of change in number density is given by
\begin{multline}
\label{eq:fix_2}
    \left(\frac{dN_{\rm c}}{dt}\right)_{\rm coag} \approx -H\sqrt{\frac{8\pi k_{\rm b} T}{m_{\rm c}}}r_{\rm c}^{2}N_{\rm c}^{2} \nu^{-1/6} \Biggl[\frac{\Gamma(\nu + 2/3,x_{\rm min})\Gamma(\nu - 1/2,x_{\rm min})}{\Gamma(\nu)^{2}} \\+  \frac{2\Gamma(\nu + 1/3,x_{\rm min})\Gamma(\nu - 1/6,x_{\rm min})}{\Gamma(\nu)^{2}} + \frac{\Gamma(\nu,x_{\rm min})\Gamma(\nu + 1/6,x_{\rm min})}{\Gamma(\nu)^2}\Biggr].
\end{multline}
Due to the reasons outlined above, in the model we only need to apply the incomplete gamma function for terms that may produce a negative or zero $\nu$ argument.
For positive arguments to the gamma function, the incomplete gamma distribution is not required to be used and the equations outlined in the main text can be applied.
This is shown in Figure \ref{fig:Kn_pop}, where, when a positive argument is given for both the incomplete gamma function and gamma function, the results are near identical. 
Therefore there is no significant loss of accuracy from using the gamma distribution for a positive $\nu$ argument than the incomplete gamma function.

For the number-weighted Knudsen number, Kn$_{N}$, presented in Eq. \eqref{eq:Kn_N}, the divergence issues are also solved by applying the incomplete gamma function rather than the gamma function, leading to
\begin{equation}
\label{eq:Kn_N_fix}
    \overline{{\rm Kn}_{N}} = \frac{\lambda_{\rm a}}{r_{\rm c}}\nu^{1/3}\frac{\Gamma(\nu - 1/3, x_{\rm min})}{\Gamma(\nu)}.
\end{equation}
As shown in Figure \ref{fig:Kn_pop}, this avoids the singularities present in the gamma distribution formulation.

At first glance, it may seem that using the incomplete gamma function does not solve the negative argument to the gamma function problem as the possibility still exists in Eqs. \eqref{eq:fix_1} and \eqref{eq:fix_2}.
However, we can utilise the recurrence relation for the incomplete gamma function
\begin{eqnarray}
    \Gamma(\nu, x_{\rm min}) = \frac{\Gamma(\nu+1,x_{\rm min}) - x_{\rm min}^{\nu}e^{-x_{\rm min}}}{\nu},
\end{eqnarray}
which ensures that a positive argument is always present when computing the incomplete gamma function.
We note that using the gamma function recurrence relation directly
\begin{eqnarray}
    \Gamma(\nu) = \frac{\Gamma(\nu+1)}{\nu},
\end{eqnarray}
would not solve this issue, as the exponential term in the incomplete gamma recurrence relation is required to properly bound the solution.
We use the numerical implementation of the upper incomplete gamma function from \citet{Press1986}, which we find is computationally stable and efficient enough for the microphysics code.

\section{Distribution weighted settling velocity}
\label{app:vf}

\begin{figure}
    \centering
    \includegraphics[width=0.95\linewidth]{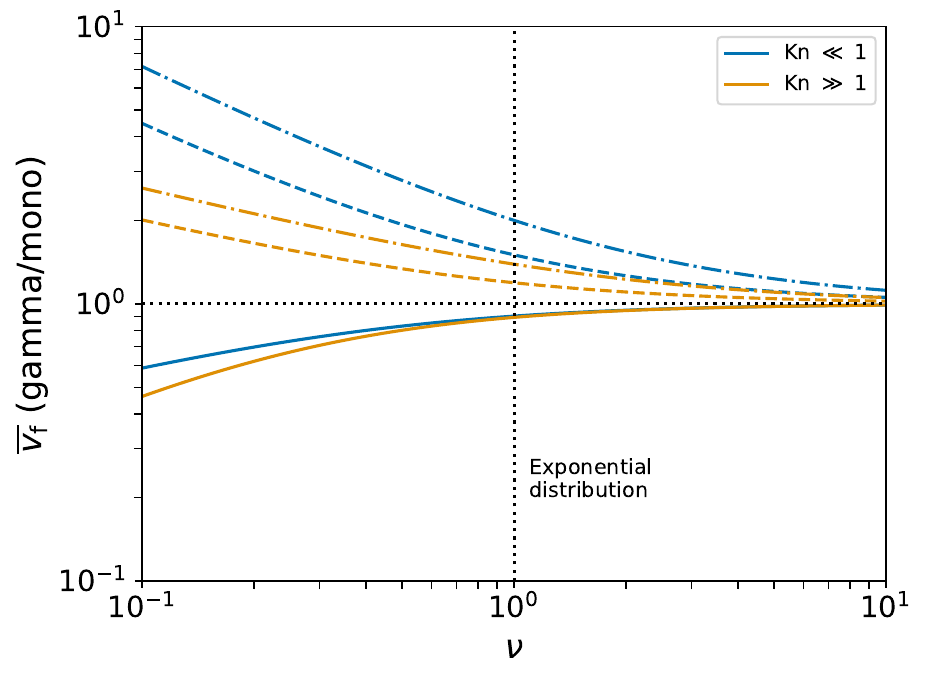}
    \caption{Relative difference between the settling velocity of the gamma distribution and the monodisperse distribution for the zeroth moment (solid lines), first moment (dashed lines) and second moment (dash-dot lines) for the Kn $\ll$ 1 regime (blue colour) and Kn $\gg$ 1 regime (orange colour).}
    \label{fig:vf}
\end{figure}

The population averaged settling velocity for a moment $k$ can be given by \citep[e.g.][]{Straka2009}
\begin{equation}
    \overline{v_{\rm f}^{(k)}} = \frac{\int_{0}^{\infty}v_{\rm f}(m)m^{k}f(m)dm}{\int_{0}^{\infty}m^{k}f(m)dm} = \frac{1}{M^{(k)}}\int_{0}^{\infty}v_{\rm f}(m)m^{k}f(m)dm,
\end{equation}
where $v_{\rm f}(m)$ [cm s$^{-1}$] is the settling velocity of a particle with mass $m$.
For the Stokes regime (Kn $\ll$ 1), we take the $v_{\rm f}$(r) expression from \citet{Ohno2017, Ohno2018}
\begin{equation}
\label{eq:vf_1}
    v_{\rm f}(r) = \frac{2\beta gr^{2}(\rho_{\rm d} - \rho_{\rm a})}{9\eta_{\rm a}}\left[1 + \left(\frac{0.45gr^{3}\rho_{\rm a}\rho_{\rm d}}{54\eta_{\rm a}^{2}}\right)^{2/5}\right]^{-5/4}.
\end{equation}
Ignoring the Cunningham slip factor and Reynolds number correction term, the population averaged velocity expression for the Stokes regime is 
\begin{multline}
    \overline{v_{\rm f,stk}^{(k)}} = \frac{qs}{M^{(k)}}\int_{0}^{\infty}m^{k+2/3}f(m)dm = \\ qs\frac{M^{(k+2/3)}}{M^{(k)}} = qs\lambda^{k + 2/3}\frac{\Gamma\left(\nu + k + 2/3\right)}{\Gamma\left(\nu + k\right)},
\end{multline}
where $q$ = $2g(\rho_{\rm d} - \rho_{\rm a})/9\eta_{\rm a}$ and $s$ = $(3/4\pi\rho_{\rm d})^{2/3}$. 
This results in the $k$ = 0, 1 and 2 population averaged settling velocities being
\begin{equation}
    \overline{v_{\rm f,stk}^{(0)}} = \frac{2g(\rho_{\rm d} - \rho_{\rm a})}{9\eta_{\rm a}}r_{\rm c}^{2}\nu^{-2/3}\frac{\Gamma\left(\nu + 2/3\right)}{\Gamma\left(\nu\right)},
\end{equation}
\begin{equation}
    \overline{v_{\rm f,stk}^{(1)}} = \frac{2g(\rho_{\rm d} - \rho_{\rm a})}{9\eta_{\rm a}}r_{\rm c}^{2}\nu^{-2/3}\frac{\Gamma\left(\nu + 5/3\right)}{\Gamma\left(\nu + 1\right)},
\end{equation}
and 
\begin{equation}
    \overline{v_{\rm f,stk}^{(2)}} = \frac{2g(\rho_{\rm d} - \rho_{\rm a})}{9\eta_{\rm a}}r_{\rm c}^{2}\nu^{-2/3}\frac{\Gamma\left(\nu + 8/3\right)}{\Gamma\left(\nu + 2\right)},
\end{equation}
respectively.
Assuming a linear form of $\beta$ (Eq. \ref{eq:b_lin}), the above equations are augmented with a moment dependent slip correction term, leading to
\begin{equation}
    \overline{v_{\rm f}^{(0)}} = \overline{v_{\rm f,stk}^{(0)}}  \Biggl[1 + A\frac{\lambda_{\rm a}}{r_{\rm c}}\nu^{1/3}\frac{\Gamma\left(\nu + 1/3\right)}{\Gamma\left(\nu+2/3\right)}\Biggr],
\end{equation}
\begin{equation}
    \overline{v_{\rm f}^{(1)}} =  \overline{v_{\rm f,stk}^{(1)}} \Biggl[1 + A\frac{\lambda_{\rm a}}{r_{\rm c}}\nu^{1/3}\frac{\Gamma\left(\nu + 4/3\right)}{\Gamma\left(\nu + 5/3\right)}\Biggr],
\end{equation}
and 
\begin{equation}
        \overline{v_{\rm f}^{(2)}} = \overline{v_{\rm f,stk}^{(2)}} \Biggl[1 + A\frac{\lambda_{\rm a}}{r_{\rm c}}\nu^{1/3}\frac{\Gamma\left(\nu + 7/3\right)}{\Gamma\left(\nu + 8/3\right)}\Biggr],
\end{equation}
respectively.

Lastly, including the Reynolds number correction factor in a fully analytical solution is more difficult due to the non-linearity of the equation.
We therefore estimate the Reynolds number factor in Eq. \eqref{eq:vf_1} as a constant, given at the respective population weighted particle size for each moments settling velocity.
We can use the gamma distribution parameters to estimate the respective population weighted mean radii.
The number-weighted mean radius, $\langle r\rangle$ [cm], is estimated from
\begin{equation}
\label{eq:r_av}
    \langle r\rangle = r_{\rm c}\nu^{-1/3}\frac{\Gamma\left(\nu + 1/3\right)}{\Gamma\left(\nu\right)},
\end{equation}
the number-weighted mass representative radius, $r_{\rm c}$ [cm], given by Eq. \eqref{eq:rc}, and the mass-weighted mass representative radius, $r_{\rm p}$ [cm], given by Eq. \eqref{eq:rp}.

The $\beta$ slip factor used in the Stokes regime is only valid between values of Kn $\approx$ 0.5-83 \citep{Kim2005}, with no guarantee that a universal asymptotic value is reached in the Kn $\gg$ 1 limit. 
In addition, we require to use a linear form of $\beta$ (Eq. \ref{eq:b_lin}) in our formulations, which, due to absence of the exponential factor,  does not coverage to a constant solution for large Kn.
We therefore include a solution for particles in the Epstein (Kn $\gg$ 1) regime.
We use the settling velocity equation from \citet{Woitke2003}
\begin{equation}
    v_{\rm f}(r) = \frac{\sqrt{\pi}g\rho_{\rm d}r}{2\rho_{\rm a}c_{\rm T}},
\end{equation}
where $c_{\rm T}$ [cm s$^{-1}$] is the atmospheric thermal velocity \citep{Woitke2003}.
Defining $q$ = $\sqrt{\pi}g\rho_{\rm d}/2\rho_{\rm a}c_{\rm T}$ and $s$ = $(3/4\pi\rho_{\rm d})^{1/3}$,
the population averaged velocity expression is 
\begin{multline}
    \overline{v_{\rm f}^{(k)}} = \frac{qs}{M^{(k)}}\int_{0}^{\infty}m^{k+1/3}f(m)dm  = \frac{qsM^{(k+1/3)}}{M^{(k)}} \\ = qs\lambda^{k + 1/3}\frac{\Gamma\left(\nu + k + 1/3\right)}{\Gamma\left(\nu + k\right)}.
\end{multline}
This results in the $k$ = 0, 1 and 2 population averaged settling velocities being
\begin{equation}
    \overline{v_{\rm f}^{(0)}} = \frac{\sqrt{\pi}g\rho_{\rm d}}{2\rho_{\rm a}c_{\rm T}}r_{\rm c}\nu^{-1/3}\frac{\Gamma\left(\nu + 1/3\right)}{\Gamma\left(\nu\right)},
\end{equation}
\begin{equation}
    \overline{v_{\rm f}^{(1)}} = \frac{\sqrt{\pi}g\rho_{\rm d}}{2\rho_{\rm a}c_{\rm T}}r_{\rm c}\nu^{-1/3}\frac{\Gamma\left(\nu + 4/3\right)}{\Gamma\left(\nu + 1\right)},
\end{equation}
and 
\begin{equation}
    \overline{v_{\rm f}^{(2)}} = \frac{\sqrt{\pi}g\rho_{\rm d}}{2\rho_{\rm a}c_{\rm T}}r_{\rm c}\nu^{-1/3}\frac{\Gamma\left(\nu + 7/3\right)}{\Gamma\left(\nu + 2\right)},
\end{equation}
respectively.

In Figure \ref{fig:vf}, we show the fractional difference between the gamma distribution population averaged settling velocity and the equivalent monodisperse settling velocity for both the Stokes and Epstein regimes.
From this, we see that the moments acquire different settling velocities to each other, with the zeroth moment settling slower than the first and second, and the second settling the fastest.
This is expected, as each moment represents a different integrated value and characteristic of the distribution which would naturally have different settling velocities \citep[e.g.][]{Woitke2020}.
For large $\nu$, the distribution becomes more monodisperse, resulting in all the moment dependent settling velocities converging on the monodisperse value.
For intermediate Knudsen number regimes (Kn $\sim$ 1) the same tanh function can be used from Paper I for each moment dependent settling velocity to interpolate between the Stokes and Epstein expressions.

\section{Turbulent shear kernel}
\label{sec:app_1}

The turbulence driven wind shear kernel is given by \citep{Saffman1956}
\begin{equation}
    K(r,r') = \left(\frac{8\pi\epsilon_{\rm d}}{15\nu_{\rm a}}\right)^{1/2}(r + r')^{3},
\end{equation}
where $\epsilon_{\rm d}$ [cm$^{2}$ s$^{-3}$] is the dissipation of turbulent kinetic energy rate of the atmosphere and $\nu_{\rm a}$ [cm$^{2}$ s$^{-1}$] the atmospheric kinematic viscosity. 
Using the \citet{Moran2022} population averaged kernel method, the resulting population averaged kernels for the zeroth and second moment are
\begin{equation}
      \overline{K(m,m')} = \frac{2K_{0}}{N_{\rm c}^{2}}\left(M^{(0)}M^{(1)} + 3M^{(2/3)}M^{(1/3)}\right),
\end{equation}
and
\begin{equation}
      \overline{K_{m}(m,m')} = \frac{2K_{0}}{\rho_{\rm c}^{2}}\left(M^{(1)}M^{(2)} + 3M^{(5/3)}M^{(4/3)}\right),
\end{equation}
respectively, where $K_{0}$ is the kinetic pre-factor
\begin{equation}
    K_{0} = \left(\frac{8\pi\epsilon_{\rm d}}{15\nu_{\rm a}}\right)^{1/2}\left(\frac{3}{4\pi\rho_{\rm d}}\right).
\end{equation}
For the gamma distribution, the number density averaged kernel is then
\begin{equation}
  \overline{K(m,m')} = 2K_{0}\lambda\left[\frac{\Gamma(\nu + 1)}{\Gamma(\nu)} + \frac{3\Gamma(\nu + 2/3)\Gamma(\nu + 1/3)}{\Gamma(\nu)^{2}}\right],
\end{equation}
and the mass-weighted kernel
\begin{equation}
  \overline{K(m,m')} = 2K_{0}\lambda\left[\frac{\Gamma(\nu + 2)}{\Gamma(\nu+1)} + \frac{3\Gamma(\nu + 5/3)\Gamma(\nu + 4/3)}{\Gamma(\nu+1)^{2}}\right],
\end{equation}
The expression for the zeroth moment is then
\begin{multline}
    \left(\frac{dN_{\rm c}}{dt}\right)_{\rm TS} \approx -\left(\frac{8\pi\epsilon_{\rm d}}{15\nu_{\rm a}}\right)^{1/2}r_{\rm c}^{3}N_{\rm c}^{2}\nu^{-1} \\ 
    \cdot\left[\frac{\Gamma(\nu + 1)}{\Gamma(\nu)} + \frac{3\Gamma(\nu + 2/3)\Gamma(\nu + 1/3)}{\Gamma(\nu)^{2}}\right],
\end{multline}
and for the second moment is
\begin{multline}
    \left(\frac{dZ_{\rm c}}{dt}\right)_{\rm TS} \approx 2\left(\frac{8\pi\epsilon_{\rm d}}{15\nu_{\rm a}}\right)^{1/2}r_{\rm c}^{3}\rho_{\rm c}^{2}\nu^{-1} \\ 
    \cdot\left[\frac{\Gamma(\nu + 2)}{\Gamma(\nu+1)} + \frac{3\Gamma(\nu + 5/3)\Gamma(\nu + 4/3)}{\Gamma(\nu+1)^{2}}\right].
\end{multline}
We can recover suitable equations for the exponential distribution in Paper I by setting $\nu$ = 1 in the above equations.

\end{document}